\documentclass[aip,jcp,reprint,amsmath,amssymb,nofootinbib]{revtex4-2}
\usepackage{graphicx,booktabs,xcolor,siunitx}
\usepackage{array,float}
\usepackage[colorlinks=true,allcolors=blue]{hyperref}
\graphicspath{{figures/}}
\newcommand{\kB}{k_{\rm B}}

\begin{document}

\title{Efficient strategies for obtaining accurate kinetics from biased simulations}

\author{Triasha Pal}
\affiliation{Department of Chemistry, New York University, New York, NY 10003}
\author{Nicodemo Mazzaferro}
\affiliation{Department of Chemistry, New York University, New York, NY 10003}
\author{Pilar Cossio}
\email{pcossio@flatironinstitute.org}
\affiliation{Center for Computational Mathematics, Flatiron Institute, New York, NY 10010}
\affiliation{Center for Computational Biology, Flatiron Institute, New York, NY 10010}
\author{Glen M. Hocky}
\email{hockyg@nyu.edu}
\affiliation{Department of Chemistry, New York University, New York, NY 10003}
\affiliation{Simons Center for Computational Physical Chemistry, New York University, New York, NY 10003}

\begin{abstract}
It has been demonstrated that one can recover the unbiased rate constants for slow conformational transitions by observing transitions driven by a bias potential added
along one or several collective variables (CVs).
However, the choice of CV, biasing scheme, biasing parameters and the rate estimator can all affect the results and it is not obvious which to choose for a given problem.
In this study, we compare accuracy versus computational cost for a previously established benchmark system, a coarse-grained protein with computationally accessible unfolding kinetics. 
We apply bias to four CVs of decreasing quality in both a time-dependent and quasi-static manner, and compare different estimators for the unfolding rate constant in terms of absolute accuracy to the estimated reference unbiased rate. 
We show that in the case of gentle biasing, the exponential average time-dependent rate (EATR) framework, which simultaneously fits the unbiased rate constant and a coordinate quality metric $\gamma$, recovers accurate kinetics in nearly all cases.
For poor CVs, estimators equivalent to assuming a good coordinate ($\gamma=1$) systematically fail regardless of biasing method or parameters.
We introduce heuristics to show when we expect the rates to be inaccurate, and identify which combinations of biasing scheme, estimator and number of runs reach a given accuracy at the lowest cost. 
\end{abstract}

\maketitle

\section{Introduction}\label{sec:intro}

A persistent challenge of studying chemical systems with molecular dynamics (MD) simulations is that most of the simulation time is spent in metastable states before observing rare transitions of interest.\cite{chandler1998barrier}
A wide range of enhanced sampling methodologies have been developed to accelerate sampling while allowing us to reweight samples and extract equilibrium information. \cite{henin2022enhanced}
A major class of these methods are those that add an additional biasing potential to the system's Hamiltonian along so-called collective variables (CVs), with the goal being to effectively allow the system to navigate the full conformational landscape by either pushing the system away from where it has already sampled or pulling the system toward regions of interest. \cite{henin2022enhanced,tuckerman2023statistical} 
Such methods include umbrella sampling,\cite{torrie1977nonphysical} metadynamics (MetaD),\cite{laio2002escaping,Barducci2008} on-the-fly probability enhanced sampling (OPES),\cite{invernizzi2020rethinking,invernizzi2020unified} adaptive biasing
force\cite{darve2008adaptive} and variationally enhanced sampling (VES).
\cite{valsson2014variational, henin2022enhanced,tuckerman2023statistical}

The same type of bias can also be used to provide information about
kinetics.~Conformational flooding,\cite{grubmuller1995predicting}  hyperdynamics,\cite{voter1997hyperdynamics,voter1997method} infrequent metadynamics
(iMetaD),\cite{tiwary2013metadynamics} OPES-flooding,\cite{ray2022rare} and variationally
optimized flooding\cite{mccarty2015variationally} all accelerate escape from a basin and allow us to recover the unbiased rate as described below. 
While a range of other path-based methods exist to compute dynamical properties from ensembles of unbiased trajectories,\cite{dellago1998transition,bolhuis2002transition} including transition interface
sampling,\cite{van2003novel} forward flux sampling,\cite{allen2009forward} weighted
ensemble\cite{huber1996weighted,zuckerman2017weighted} and
milestoning,\cite{elber2020milestoning} these methods accelerate sampling via different mechanisms from adding bias.
In our work, we have focused our attention on increasing the effectiveness of kinetics estimation from added bias along good or poor CVs. In this work, we concentrate on the accuracy-cost trade-off within that branch of approaches.

Recovering kinetics from biased simulations rests on the idea that the additional bias effectively destabilizes the starting state or equivalently lowers the barrier to the transition of interest. 
It is assumed that if kinetics are dominated by crossing a high barrier and bias is added in such a way that it does not pollute the transition barrier, then transitions occur via the same mechanism as in the unbiased case.
OPES- and VES-flooding try to avoid biasing transitions by applying a quasi-static bias capped below the
barrier and potentially excluded outside a specified region,\cite{ray2022rare,mccarty2015variationally} while 
iMetaD and similar methods add a time dependent bias that changes infrequently such that a fleeting number of transitions occur between changes in the biasing potential.\cite{tiwary2013metadynamics}

Under the assumptions where Kramers' theory would hold, a bias $V$ added along a perfect reaction coordinate accelerates transitions on average by the acceleration factor $\alpha=\langle e^{\beta V}\rangle$. \cite{grubmuller1995predicting,voter1997hyperdynamics,tiwary2013metadynamics}
Due to the exponential scaling in energy of the rate constant, for problems dominated by a very high barrier, it is possible to achieve enormous savings in computational effort.

All of these methods assume that the CV is a good reaction coordinate for the transition, or equivalently that all bias added along the CV goes into accelerating the transition of interest.
We know this assumption will fail in all but idealized cases. This issue is often exacerbated by the fact that a CV must be chosen before the transition mechanism is understood.
In the case of a non-ideal CV,  the estimated acceleration of dynamics is often overestimated because bias accumulates along irrelevant degrees of freedom that do not facilitate the transition. 
Consequently, the reweighted transition times are artificially inflated, leading to an underestimated rate constant.
One approach to this problem is to iteratively improve the CV used with data from earlier biased trajectories, and it is possible to use some of this kinetic estimation to help in guiding that iteration.\cite{wang2021state,mazzaferro2025using}

Alternatively, we can try to simultaneously predict the unbiased kinetics while measuring the quality of the selected CV.\cite{palacio2022transition}
With our exponential average time-dependent rate (EATR) method,
\cite{mazzaferro2024good} we showed that we could extract accurate rate constants from an ensemble of trajectories experiencing a time-dependent bias using a second parameter $\gamma$.
In a more recent work, we introduced EATR-flooding that extends the same concept to static biases by fitting $\gamma$ using sets of runs experiencing different levels of bias.\cite{mazzaferro2026stepping}
For a perfect CV ($\gamma = 1$), EATR recovers the usual acceleration factor.
For bad CVs, EATR applied to MetaD is often able to recover accurate kinetics even when bias is added at a relatively frequent pace. 

For each choice of biasing method, there are a number of parameters we must choose that each could have an effect on efficiency and accuracy. 
For example, iMetaD results depend on the hill addition pace, the hill height, and the bias factor, whereas for the quasistatic OPES-flooding or VES-flooding methods the most important parameter is the maximum level of bias added. 
For a new system and CV, it is not obvious which choice of method or parameters will lead to the best result. 
Aggressive biasing shortens the average observed transition time, but can perturb the transition region or drive the system so hard that the reaction is not thermally driven. 
On the other hand, slow biasing or low levels of bias result in inaccurate results if only a few transitions can be observed and in the case of the EATR estimator, also makes it harder to extract the unknown $\gamma$ parameter.

The practical question investigated in this work is which scheme,
parameters and estimator give an accurate rate for the least simulated time.
We do so using a previously investigated coarse-grained model for protein folding, chosen because its unbiased unfolding rate can be measured
directly and because CVs of very different quality are easy to
define.\cite{dudko2011locating,mazzaferro2024good}
We assess our ability to extract accurate transition times from folded to unfolded for four CVs and using multiple time-dependent and quasi-static biasing approaches. 
We show that all biasing approaches can achieve equal accuracy, with time-dependent biasing reaching higher accuracy at lower cost when combined with EATR. However, the quasi-static approaches have more consistent accuracy and it is easier to control the error introduced by the added bias.
For bad CVs, we show that EATR is always necessary to correct the systematic error introduced by the biasing.
Using our data set, we give heuristics for how fast and how much bias one can add and still achieve accurate results, which lead to guidelines for how to structure the biased simulations for a new system.

The rest of the paper is organized as follows:
Section~\ref{sec:methods} defines the estimators and benchmark; Sec.~\ref{sec:results}
compares their accuracy, data requirements and cost, then identifies the regimes where rate estimation succeeds and fails; Sec.~\ref{sec:conclusions} gives our general conclusions; and finally Sec.~\ref{sec:compdetails} gives all computational details for the large data set. All code, data and results are available in a repository accompanying this manuscript (see Data Availability).

\section{Theory and Approach}\label{sec:methods}
In this section, we give the most important details of the theoretical frameworks used in this study and introduce all key notation, with more in depth description of our simulations provided in the Computational Details Sec.~\ref{sec:compdetails}.

\subsection{Computing free energies by collective variable biasing}\label{sec:fes}
We use MD simulations at temperature $T$ to sample configurations $\mathbf x$ from the canonical distribution of a system described by a
Hamiltonian $H(\mathbf x, \mathbf p) = K(\mathbf p) + U(\mathbf x)$.
In this case, we expect to sample configurations with probability 
$\propto e^{-\beta U(\mathbf x)}$, with $\beta = 1/\kB T$.\cite{tuckerman2023statistical}
Integrating out every degree of freedom except a collective variable $s(\mathbf x)$
defines the free energy surface (FES), or potential of mean force (PMF),
\begin{equation}
  F(s) = -\frac{1}{\beta}\ln\int d\mathbf x\,\delta\big(s - s(\mathbf x)\big)\,
  e^{-\beta U(\mathbf x)} + {\rm const},
  \label{eq:fes}
\end{equation}
so that $s$ is also Boltzmann distributed as $P(s) \propto e^{-\beta F(s)}$. A barrier of many $\kB T$
on the FES makes transitions between its basins rare, and $P(s)$ hard to sample.

CV-based enhanced-sampling methods add a bias $V(s)$ to the Hamiltonian so the biased system is represented by 
$H + V(s(\mathbf x))$, which changes the equilibrium distribution of the CV to
$P_V(s) \propto e^{-\beta[F(s) + V(s)]}$; the bias depends on $\mathbf x$ only through
$s$, so it is constant over the $\delta$-function in Eq.~\eqref{eq:fes} and factors out of
the integral. If the bias is static and known, the FES
follows from the biased histogram, $F(s) = -\beta^{-1}\ln P_V(s) - V(s) + {\rm const}$,
as in umbrella sampling;\cite{torrie1977nonphysical} a good static bias must, however,
be designed in advance. 
Adaptive methods instead build a time-dependent bias $V(s,t)$ from the trajectory itself, so that $V$ approaches a known function of $F$ and the FES
can be read from the bias or recovered by reweighting.

In this work, we study three adaptive schemes, which differ in how the bias is built and whether bias
continuously grows throughout a simulation.
We write the equations for one CV; they extend directly to higher dimensions.

\emph{Metadynamics.} MetaD builds the bias as a sum of repulsive Gaussian hills
deposited at the current value of the CV every $\tau$ time steps (\texttt{PACE} in PLUMED),\cite{laio2002escaping}
\begin{equation}
  V(s,t) = \sum_{k\,:\,k\tau < t} h_k\,
  \exp\!\left[-\frac{(s - s_k)^2}{2\sigma^2}\right],
  \label{eq:metad}
\end{equation}
with $s_k$ the CV at the $k$th deposition and $\sigma$ the hill width. In the original,
untempered form every hill has the same height, $h_k = h_0$. The bias fills the minima
of $F$ one after the other, and at long times it fluctuates around $-F(s)$ rather than
converging, with fluctuations set by the hill height and the deposition
pace.\cite{laio2002escaping,bussi2006equilibrium} Well-tempered MetaD (WT-MetaD) scales each
hill down by the bias already present at the current position in CV space,
$h_k = h_0 e^{-V(s_k,t_k)/((b-1)\kB T)}$, with bias factor $b > 1$.\cite{Barducci2008} The
bias then converges smoothly to $V(s) \to -(1 - 1/b)F(s)$, so that $s$ is sampled from
$P(s)^{1/b}$, a distribution with barriers lowered by the factor $b$. Untempered
MetaD and unbiased sampling arise in the limits of $b \to \infty$ and
$b \to 1$.\cite{Barducci2008}

\emph{On-the-fly Probability Enhanced Sampling.} OPES builds the bias from an on-the-fly estimate of the probability
distribution of the CV rather than from a sum of hills,\cite{invernizzi2020rethinking}
\begin{equation}
  V_t(s) = \left(1 - \frac{1}{b}\right)\frac{1}{\beta}
  \ln\!\left(\frac{\tilde P_t(s)}{Z_t} + \epsilon\right),
  \label{eq:opes}
\end{equation}
where $\epsilon = e^{-\beta\Delta E/(1-1/b)}$ caps the bias near the \texttt{BARRIER}
parameter $\Delta E$.\cite{invernizzi2020rethinking} In the flooding regime,\cite{ray2022rare} kernels
are not deposited in an excluded region that contains the transition state, and
$\Delta E$ sets the fill of the reactant basin. 

\emph{Variationally Enhanced Sampling.} Developed prior to OPES, VES uses a similar but distinct strategy to obtain the applied bias by minimizing a convex functional whose
minimum samples a target distribution
$p_{\rm tg}$,\cite{valsson2014variational,valsson2020variationally}
\begin{equation}
  \Omega[V] = \frac{1}{\beta}\ln\frac{\int ds\, e^{-\beta[F(s)+V(s)]}}{\int ds\, e^{-\beta F(s)}}
  + \int ds\, p_{\rm tg}(s)V(s),
  \label{eq:ves}
\end{equation}
with the bias expanded in basis functions, $V(s;\mathbf c) = \sum_i c_i f_i(s)$, and the
coefficients optimized by averaged stochastic gradient descent.\cite{bach2013non} 

\begin{figure*}[ht]\centering
\includegraphics[width=6.69in]{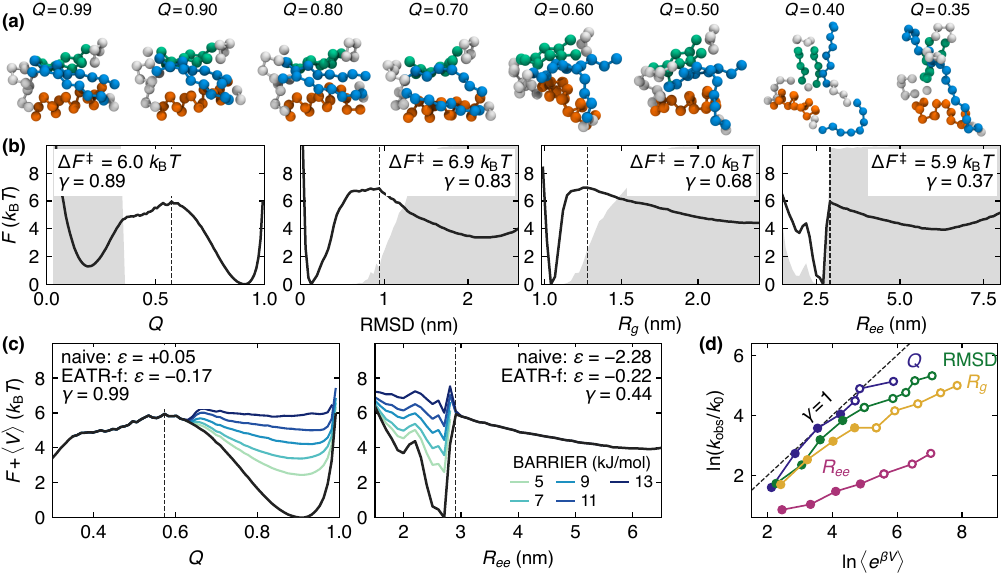}
\caption{The benchmark system, its free energies along different CVs, and analysis of CV quality. (a) The C$_\alpha$ G\=o-like model of protein G at a ladder of $Q$ values; each snapshot is a frame of one unbiased first-passage run, and the one at $Q = 0.35$ precedes commitment to the unfolded state by under 1~ns. The N-terminal hairpin ($\beta_1$, $\beta_2$) is shown in blue, the helix in vermilion, the C-terminal hairpin ($\beta_3$, $\beta_4$) in green and the loops in gray. (b) Unbiased PMFs,
$F(s) = -\kB T\ln P(s)$, along the four biased CVs were computed from histograms (100 bins) of a
120~$\mu$s unbiased trajectory sampled every 100~ps, with $F = 0$ at the folded
minimum. The gray fill shows the fraction of frames in each bin that are unfolded ($Q < 0.35$),
and the dashed line marks the apparent barrier, the PMF maximum between the folded and unfolded minima.
The value of $\gamma$ given in each panel is the median EATR-flooding efficiency of the OPES bias levels over the eight paces.
(c) The biased landscape $F + \langle V\rangle$ is shown for OPES-flooding at pace 1000 on $Q$ and $R_{ee}$,
for the five \texttt{BARRIER} values $\Delta E$. The average bias
$\langle V\rangle(s)$ is the OPES bias averaged in each CV bin over the second half of
each of the 50 runs of a set, when the bias has converged, shifted by $\Delta E$ so that
it vanishes where no kernels were deposited; bins with fewer than 20 samples are
omitted. The annotations give the error $\varepsilon = \ln k_{\rm est} - \ln k_0$ of the
naive and EATR-flooding estimates from those five levels. (d) The observed escape rate of every
OPES set at pace 1000 is plotted against its acceleration factor, with $k_0$ the unbiased
reference rate; on the dashed diagonal the naive estimate is exact. Open symbols denote levels
above 13~kJ/mol, which were run with the same parameters to extend these curves but were not used for rate calculations below.}\label{fig:system}
\end{figure*}

\subsection{Computing kinetics by collective variable biasing}\label{sec:rates}
Each of the three biasing schemes above can be adapted to compute kinetics, where the speedup gained by biasing can be computed for particular choices of parameters.
The goal is to make it such that the bias fills the reactant basin while leaving the transition region unperturbed. 
Under this condition, and where Kramers' theory applies, the unbiased mean first-passage time $t$ follows from
the biased time $t_f$ through the acceleration
factor\cite{voter1997hyperdynamics,tiwary2013metadynamics}
\begin{equation}
  \alpha = \frac{t}{t_f} = \left\langle e^{\beta V(s)} \right\rangle_{V},
  \label{eq:accel}
\end{equation}
with the average taken over the biased trajectory in the reactant basin. 
Traditionally, one runs many biased simulations and then rescales time in each one to produce a separate acceleration factor, $\alpha_i$, depending on the particular biased escape trajectory that is sampled. 

We can write all of the estimators we will use to extract unbiased rate constants in one form by considering a generalized rate scaling function $f(t)$ for each trajectory. \cite{mazzaferro2024good}

Run $i$ experiences a bias $V_i(t)$, and the
survival function of a collection of runs is given by
\begin{equation}
  S(t) = \exp\!\left[-k_0\int_0^t f(t')\,dt'\right],
  \label{eq:survival}
\end{equation}
with $k_0$ the unbiased rate. \cite{mazzaferro2024good}

Below we use an ``overbar'' to indicate an average across simulations at the same time. 
For iMetaD, $f = \overline{e^{\beta V}}$, equivalent to rescaling each transition time by
$\alpha_i = t_i^{-1}\int_0^{t_i} e^{\beta V_i}dt$ and fitting the empirical cumulative
distribution of the rescaled times $\alpha_i t_i$ to $1 - e^{-k_0 t}$. 
Previously it was proposed that the Kolmogorov--Smirnov (KS) test of that distribution against an
exponential \cite{salvalaglio2014assessing} can be used to check the assumption
of an unbiased transition region. EATR uses $f(t) = \overline{e^{\beta\gamma V(t)}}$,
averaged over the runs still in the basin at $t$, and fits $(k_0, \gamma)$ to the
empirical cumulative distribution of the observed times.\cite{mazzaferro2024good}
The iMetaD acceleration factor of Eq.~\eqref{eq:accel} is the special case where $\gamma = 1$.

For a static bias the survival function is
exponential, and $k_0$ and $\gamma$ cannot be separated from one set of runs. Naive
flooding ($\gamma = 1$) takes $\ln k_0 = \ln k_{\rm obs} - \ln\langle e^{\beta V}\rangle$,
with $k_{\rm obs}$ the maximum-likelihood rate of the (possibly censored) transition
times of the set. 

In EATR-flooding \cite{mazzaferro2026stepping} we introduced the idea of instead taking several sets
$j$ of different bias amplitude,
\begin{equation}
  \ln k_{{\rm est},j}(\gamma) = \ln k_{{\rm obs},j} - \ln\langle e^{\beta\gamma V}\rangle_j,
  \label{eq:eatrf}
\end{equation}
and choose $\gamma^\ast = \arg\min_\gamma {\rm Var}_j[\ln k_{{\rm est},j}(\gamma)]$; we then
report $\ln k_0 = \langle \ln k_{{\rm est},j}(\gamma^\ast)\rangle_j$. For a good CV,
$\ln k_{\rm obs}$ rises one-for-one with $\ln\langle e^{\beta V}\rangle$; its slope is
a good approximation to $\gamma$, an example of which will be seen in Fig.~\ref{fig:system} as described next.\cite{mazzaferro2026stepping}

In MetaD, the hill deposition pace controls how quickly the bias grows.
For small deposition time $\tau$, energy rises rapidly but
leaves little time for relaxation between hills, with the system ``surfing'' along the bias at particularly low paces. 
Kinetic calculations therefore deposit
infrequently, ideally avoiding bias in the transition region before
escape.\cite{tiwary2013metadynamics}
Frequency-adaptive MetaD instead starts from a short deposition interval and lengthens it as the acceleration factor grows, so that the bias fills the basin quickly and then slows before it can reach the transition state.\cite{Wang2018}
The bias factor, initial height $h_0$ and pace together
set the fill at escape, an approximate formula for which is given in the supplementary material, Sec.~\ref{sec:s_design}. In our systematic studies, we vary the
pace over three decades because it most directly trades cost against accuracy
(Sec.~\ref{sec:compdetails}).

For OPES or VES, over-biasing is instead controlled by setting the maximum bias level; this is done by setting the maximum level of bias below what one has previously estimated as the transition barrier in the particular CV space being biased. 
In the case of VES, a uniform target is combined with a bias cutoff $F_c$ above which the bias may not grow,\cite{mccarty2015variationally} and the converged bias is then applied as a
static potential. 
VES-flooding thus separates bias construction from production, so that every escape trajectory experiences exactly the same time-independent bias.
In OPES flooding where the maximum bias level $\Delta E$ is deliberately below the transition barrier, the average bias within the well converges quickly and then becomes effectively
static.

In VES-flooding, the bias can by construction be set to zero outside of a particular defined region, allowing the strategy to avoid applying bias during the transition.
In OPES, it is also possible to define an excluded region outside of which the bias is not applied.
This can be a useful strategy, but can also make simulations take longer given that no bias helps the system reach the barrier within that excluded region.
Our previous and present results indicate that accurate rates can also be obtained with EATR-flooding without including an excluded region.\cite{mazzaferro2026stepping} 

\subsection{Benchmark system, collective variables and reference rate}\label{sec:system}
A good benchmark system for rate strategies needs a transition whose rate is known independently,
coordinates of different and hypothesized quality, and a cost per run low enough to repeat
every protocol hundreds of times. We previously adopted the unfolding of the B1 domain of protein G in a
C$_\alpha$ G\=o-like model\cite{karanicolas2002origins,karanicolas2003improved,takada2019go}
 to satisfy these criteria.
 The domain is a two-state folder whose folding mechanism, and which coordinates do and do not describe it, had been characterized in
detail (Fig.~\ref{fig:system}a). \cite{mazzaferro2024good,mazzaferro2026stepping,dudko2011locating,graham2011force} 
Moreover, its unfolding is fast enough in the coarse-grained model to be sampled directly.

Four CVs are biased. The fraction of native contacts,
\begin{equation}
  Q = \frac{1}{N_c}\sum_{(i,j)} \frac{1}{1 + e^{\beta_Q\left(r_{ij} - \lambda r_{ij}^0\right)}},
  \label{eq:q}
\end{equation}
sums over the $N_c$ native contacts, with $r_{ij}^0$ the distance in the folded
structure, $\beta_Q = 50$~nm$^{-1}$ and $\lambda = 1.2$. The other three are the
C$_\alpha$ root-mean-square deviation (RMSD) from the folded structure after optimal
alignment, the radius of gyration $R_g$ of the C$_\alpha$ beads, and the end-to-end
distance $R_{ee}$ between the terminal beads. Their unbiased PMFs, computed from a
120~$\mu$s unbiased trajectory (Fig.~\ref{fig:system}), give basin widths
$\sigma_A = 0.050$, 0.066~nm, 0.014~nm and 0.152~nm and apparent barriers of 5.95, 6.90,
6.96 and 5.92~$\kB T$.

The reference rate comes from $N = 400$ independent unbiased runs started in the folded
state. Every run unfolded, so the maximum-likelihood estimate of an exponential escape
rate is $k_0 = N/\sum_i t_i$, with $t_i$ the first-passage times. Its uncertainty is the
standard deviation over 2000 bootstrap resamples of the runs. This gives
$k_0 = 1.54~\mu{\rm s}^{-1}$, $\ln(k_0/\mu{\rm s}^{-1}) = 0.431 \pm 0.050$, consistent with
earlier estimates on this model,\cite{mazzaferro2024good,mazzaferro2026stepping} at a
total cost of 260~$\mu$s of simulation. The mean first-passage time, $1/k_0$, is
0.65~$\mu$s. 
The first-passage times are
exponential (Kolmogorov--Smirnov statistic $D = 0.029$, below the 5\% critical value of 0.068 for $N = 400$), as expected for a rare event. We measure our errors using
$\varepsilon = \ln k_{\rm est} - \ln k_0$ in natural-log (ln) units. Details about the biased
simulations---a grid of WT-MetaD protocols and sets of OPES- and VES-flooding bias
levels---are specified in Sec.~\ref{sec:compdetails}.
All simulations use LAMMPS \cite{thompson2022lammps} with PLUMED; \cite{plumed2019promoting}
each run starts in the folded state and ends on unfolding, defined by a fraction of native
contacts $Q < 0.35$.

\begin{figure*}[ht]\centering
\includegraphics[width=6.69in]{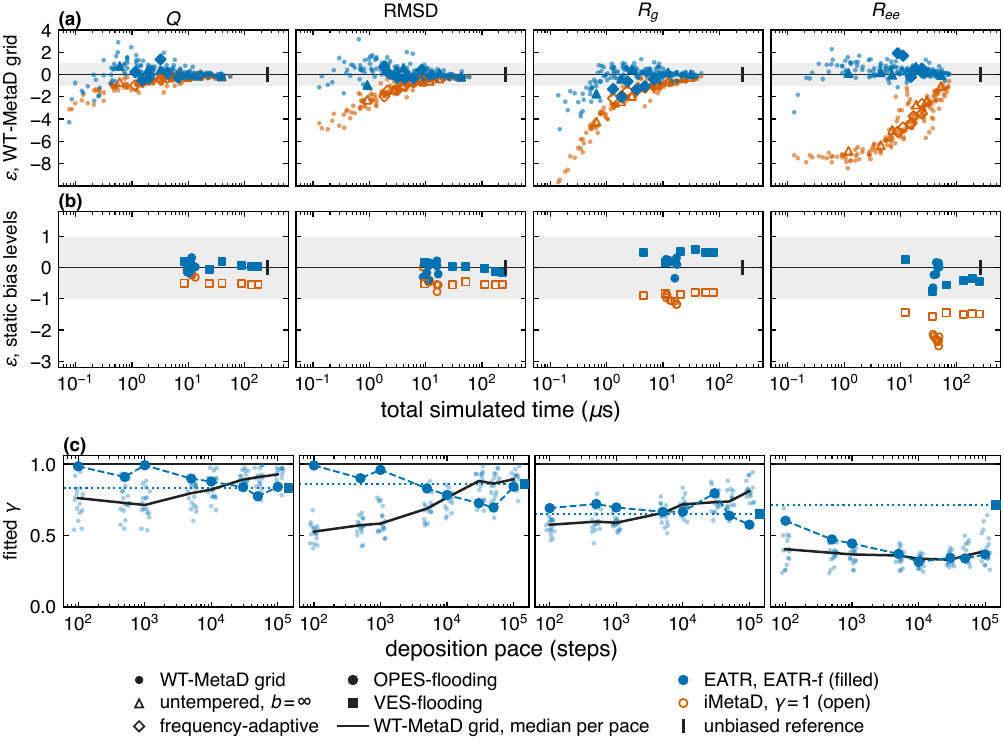}
\caption{Signed error in the $\ln$ of the estimated rate constant, $\varepsilon = \ln k_{\rm est} - \ln k_0$, against total simulated time for every data set, and the biasing efficiency $\gamma$ fitted to the same data. For either estimator, $\varepsilon > 0$ means the estimated rate is too fast, $k_{\rm est} > k_0$, and $\varepsilon < 0$ that it is too slow, $k_{\rm est} < k_0$. The columns are
the four CVs. The marker shape indicates the protocol, and the color/fill indicates the estimator: blue and filled where $\gamma$
is fitted (EATR, EATR-flooding), vermilion and open where it is taken to be 1 (iMetaD, naive
flooding). Individual WT-MetaD results are represented as small filled dots in the same two colors. (a) Analysis is applied to WT-MetaD simulations with  estimators applied on identical data, one point per condition
(50 or 200 runs, Sec.~\ref{sec:compdetails}); triangles are untempered MetaD, and diamonds frequency-adaptive MetaD, with six frequency thresholds at $b = 10$ and $h_0 = 0.19~\kB T$ (Sec.~\ref{sec:accuracy}).
(b) Analysis applied with static and quasi-static biasing. For OPES-flooding, we show one point per pace (five \texttt{BARRIER} values, 50 runs
each), and for VES cutoff flooding, one point per number of runs per cutoff. 
In (a) and (b), the black bar shows the unbiased
reference (from 400 runs) and the gray band shows one $\ln$ unit. (c) The efficiency parameter $\gamma$
obtained by EATR versus deposition pace. Points show every WT-MetaD condition with bias factor
5--20 (200 runs each), whereas the line shows the median for each pace. Circles show EATR-flooding over the five
OPES levels at each pace. The VES bias is static and has no pace, so its EATR-flooding fit
over the six cutoffs is drawn as a dotted level with a square at the right edge.}\label{fig:cost}
\end{figure*}

\section{Results and Discussion}\label{sec:results}
\subsection{Overview of the benchmark system}\label{sec:overview}

We first compute the PMF for each of the CVs, which we previously showed range from a near-ideal reaction coordinate to a poor one
(Fig.~\ref{fig:system}b). Along $Q$ the folded and unfolded basins are separated at
the barrier. Along RMSD and $R_g$ unfolded frames begin to populate the barrier region.
Along $R_{ee}$ the barrier top lies inside the unfolded ensemble (98\% of the frames in
the barrier bin are unfolded), and the folded basin overlaps configurations on the way
to unfolding. A static OPES bias shows the consequence of this for $Q$ and $R_{ee}$ (Fig.~\ref{fig:system}c; the
corresponding landscapes for every protocol are in Fig.~\ref{fig:s_overlays}). On
$Q$ the bias fills the folded basin to a level plateau at or below the barrier top (within 0.3~$\kB T$ of it for the two highest levels) and leaves the
barrier untouched, and both the naive estimate and EATR-flooding recover the
rate. 
On $R_{ee}$ the bias raises the landscape just inside the barrier by 1--3~$\kB T$,
so part of the biasing is spent in a location that does not promote escape; the
naive estimate of the rate constant is low by 2.3 $\ln$ units, while EATR-flooding, with $\gamma = 0.44$,
is within 0.22. 
In Figure~\ref{fig:system}d, we demonstrate how our EATR-slope analysis can graphically reveal a fraction of the expected acceleration that is wasted; for $Q$, the observed rate rises nearly identically with the acceleration factor showing efficient biasing, whereas along
$R_{ee}$ it rises only at a slope of about 0.4. 


\subsection{EATR and EATR-flooding estimate accurate rate constants for protocols spanning a wide range of computational costs}\label{sec:accuracy}

\begin{figure}[t]\centering
\includegraphics[width=\columnwidth]{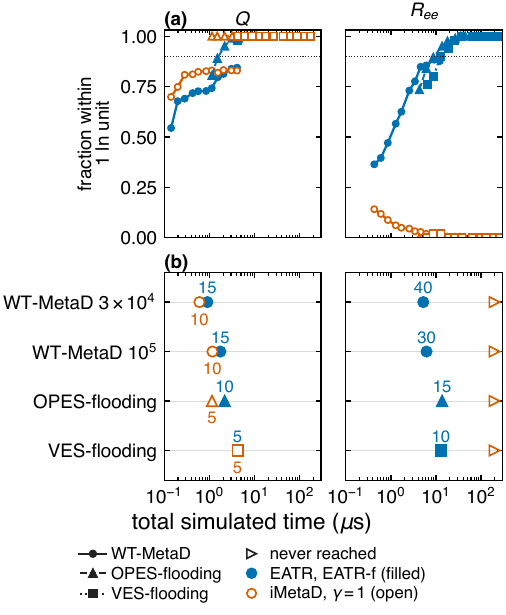}
\caption{How much data each strategy needs, for $Q$ and $R_{ee}$. Shape and line style
give the protocol, and color and fill give the estimator as in Fig.~\ref{fig:cost}, with blue and
filled symbols where $\gamma$ was fitted and vermilion and open symbols where $\gamma = 1$. (a) The fraction of
random subsets of $n$ runs within one $\ln$ unit of the reference is plotted against the median cost of
a subset, and the dotted line marks nine in ten. WT-MetaD is the mean over grid conditions (bias factors
2--20), OPES is the mean over paces, and VES is the single sweep. (b) The minimum data is the smallest
cost at which that fraction reaches nine in ten and stays there at every larger size. The
number beside each symbol is how many runs that is (per condition for WT-MetaD and per bias
level for the flooding protocols), and an open arrow at the right edge marks a case that
never reaches the criterion. Subset sizes form a ladder with steps of about a factor of 1.5
(5, 7, 10, 15, 20, 30, 40, 60, 80, 120 and 160 runs per condition for WT-MetaD; 5 to 40 per
level for OPES and 5 to 150 per cutoff for VES in similar steps, plus the full set), so each
minimum is resolved to that factor. WT-MetaD is shown at two slow paces,
$3\times10^4$ and $10^5$ steps (300~ps and 1~ns), which correspond to 0.5--3.3 and 1.6--11 relaxation
times of the CVs (Sec.~\ref{sec:works}).
At $10^3$ steps and faster, neither estimator
reaches the criterion for any CV. Results for all four CVs are shown in Fig.~\ref{fig:s_mindata} of the
supplementary material.}\label{fig:mindata}
\end{figure}

We define the cost of a rate constant estimate, whether obtained by rescaling time with computed acceleration factors (the ``naive'' estimator) or by EATR, as the total MD time used to make that estimate.
Because runs are terminated when a transition is observed, strongly biased simulations may have very low total cost even if many trajectories are used. However, this over-biasing can lead to systematic errors which cannot be corrected by increasing the number of runs.

Figure~\ref{fig:cost} summarizes broad parameter sweeps over MetaD, OPES and VES for all
four CVs; Figs.~\ref{fig:s_grid_Q}--\ref{fig:s_grid_Ree} of the supplementary material show every individual condition, and Fig.~\ref{fig:cost}c, together with Fig.~\ref{fig:s_gamma} and Table~\ref{tab:s17} of the supplementary material, collect the $\gamma$ values behind the EATR estimates.

For the growing biases (Fig.~\ref{fig:cost}a), EATR is within one $\ln$ unit on 84--88\%
of the WT-MetaD grid for every CV. In contrast, iMetaD falls from 82\% for $Q$ to 58\% for
RMSD, 38\% for $R_g$ and 1\% for $R_{ee}$. Untempered MetaD shows the same pattern: EATR
succeeds in 23 of 24 conditions, whereas iMetaD is low by a median 3.8 $\ln$ units for
$R_{ee}$. Thus, untempered deposition fits into the data in line with the rapidly growing, $b\to\infty$ limit.

However, EATR does not rescue estimates from deposition that is too fast. At paces of 1--10~ps, escape becomes driven by the bias and both estimates scatter by more than one $\ln$  unit.
As described in a later section, the relevant quantity for failure is not the pace but the pace relative to CV relaxation time
(Sec.~\ref{sec:works}).

Frequency-adaptive MetaD\cite{Wang2018} (diamonds in Fig.~\ref{fig:cost}a) is designed to mitigate this problem by starting from a short deposition interval and lengthening it as the acceleration grows. 
However, we find that this strategy does not increase accuracy at fixed cost, with its results falling along the WT-MetaD data with fixed pace.
Nor does it reach the lowest costs: its most
aggressive setting costs five to seven times more than the fastest fixed pace of our grid (100 steps) for $Q$, RMSD and $R_g$, and 26 times more for $R_{ee}$, although that is a pace at which, as noted above, neither estimator is reliable.
What it can provide is robustness, because the
iMetaD error degrades smoothly with the acceleration threshold, so it can still be useful when the proper pace has not been estimated in advance (Fig.~\ref{fig:s_fametad}).

Static biases have a similar overall message (Fig.~\ref{fig:cost}b). Across the five OPES levels,
EATR-flooding is within 0.7 $\ln$ units for every CV and pace. Naive estimates remain low,
$\gamma=1$ effectively introducing an error of 
$(1-\gamma)\ln\langle e^{\beta V}\rangle$.
However, this error is more controlled than in the case of growing bias, which we can see by comparing the results for $R_{ee}$, where the naive error is $\approx-2.3$ $\ln$  units for OPES rather than
$< -6$ for many iMetaD conditions.
VES results, generated via a truly static and tailored potential, perform similarly to OPES-flooding, although the naive estimate is slightly better in the case of $R_{ee}$.

Across our data sets, the fitted $\gamma$ depends most strongly on the property of the coordinate and not the individual methods' parameters
(Fig.~\ref{fig:cost}c and Fig.~\ref{fig:s_gamma}). 
At deposition paces of $10^4$ steps and slower, the median $\gamma$
is 0.90 for $Q$, 0.85 for RMSD, 0.74 for $R_g$ and 0.35 for $R_{ee}$.
At these paces, changing the MetaD bias factor from 5 to 20 moves the median $\gamma$ by at most 0.04 for any CV.
As observed in, \cite{mazzaferro2024good} faster deposition generally lowers the effective $\gamma$ as more energy is wasted, although the average trend for the worst CV,  $R_{ee}$, is mostly flat 
(See also Table~\ref{tab:s17}).
$\gamma$ values extracted from EATR-flooding with OPES are consistent with very slow deposition of MetaD, and are higher than the MetaD values at paces up to 5000 steps. The efficiency of OPES decreases for very slow deposition because it takes longer to converge to the static bias used in the estimates. 

Could the failures of the naive estimators have been detected without the reference
rate? The usual check is a Kolmogorov--Smirnov (KS) test of the rescaled escape times
against an exponential (Sec.~\ref{sec:rates}). For a growing bias, we find that the KS test is
sensitive but not specific (Figs.~\ref{fig:s_ks}--\ref{fig:s_cdf_imetad}).
Of the 352 iMetaD conditions that disagree with the reference value by more than one $\ln$ unit, 334 fail with $p < 0.05$; however,  28--44\% of the accurate estimates for $Q$, RMSD and $R_g$ also fail the KS test. The mismatch between the KS test and accuracy decreases as the pace is raised from 100 to $10^5$ steps, at which point it becomes a relatively predictive metric.
EATR fails
the test only at paces up to 1000 steps, where $\gamma$ is depressed by fast deposition, and
never at slower paces, whether or not the barrier is overfilled, so the test reveals one of the two routes that lead to inaccurate EATR results
(Sec.~\ref{sec:works}). 
For a static bias, the rescaled
escape times remain exponential, so only 4 of the 82 OPES and VES levels whose naive estimate is
off by more than one $\ln$ unit fail the test; the other 78 pass despite being inaccurate.
Fitting $\gamma$, or reading the slope of $\ln k_{\rm obs}$ against
$\ln\langle e^{\beta V}\rangle$ (Fig.~\ref{fig:system}d), remains the only diagnostic for a
static bias.

\subsection{The number of simulations required for accurate kinetics}\label{sec:mindata}
As a general rule, benchmark studies on kinetics methods are able to use a large number of simulations to converge rate constant estimates, but much fewer runs are used in production calculations due to practical limitations of available simulation time. 
We therefore wish to ask, at least for this simplified data set, how few simulations can we afford and still obtain accurate results. 

To measure this data requirement, we refit random subsets of $n$ runs. We define the minimum as the
smallest cost at which nine subsets in ten fall within one $\ln$  unit of the reference and
continue to do so at every larger size. 

In Fig.~\ref{fig:mindata}a, we show how this fraction increases as additional data is provided to the estimators for a good and bad CV (with all four shown in Fig.~\ref{fig:s_mindata}). 
In Fig.~\ref{fig:mindata}b, we list the minimum number of runs (for the discrete sizes tested) at which our accuracy criterion is met. 

As also indicated in Fig.~\ref{fig:cost}, for a good CV, the error in EATR can actually be higher at fixed cost  than the naive estimator since noise is introduced by trying to fit a second parameter. 
For the protocols shown in Fig.~\ref{fig:mindata}b, we needed up to twice the number of runs to reach the same level of accuracy when fitting a high-quality CV with EATR or EATR-flooding.
Yet, for a bad CV, the naive estimator never reaches anywhere close to this accuracy threshold for any of the protocols, and with EATR or EATR-flooding we are able to recover accurate kinetics robustly with 10s of separate simulations.

This subsampling analysis is implemented in our EATR analysis code, and checking whether smaller sets of data give consistent values of $\gamma$ and estimated rate is an important and easy self-consistency check that can be performed in most cases except when data is extremely limited due to high computational expense. 

\subsection{The accuracy-cost trade off across methods}\label{sec:front}
\begin{figure}[t]\centering
\includegraphics[width=\columnwidth]{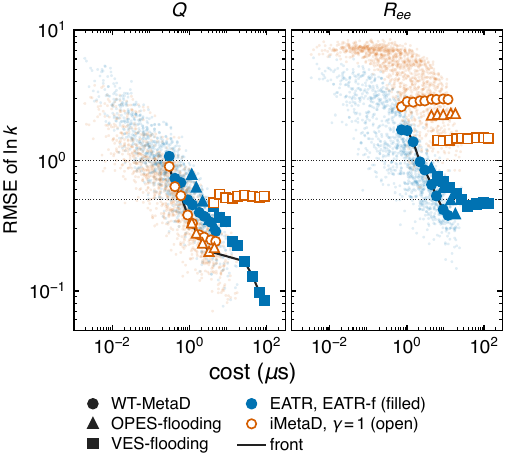}
\caption{Error--cost front by strategy for $Q$ and $R_{ee}$. Large symbols show the typical error of
$\ln k$ (root-mean-square over independent subsets) against the median cost of the production
runs, for WT-MetaD on the gentle conditions of the grid, selected in advance by the rule of
the supplementary material, and for the flooding protocols over all paces, with symbols as in
Fig.~\ref{fig:mindata}. The solid line shows the front, the strategies that have the lowest typical error for a particular cost. Faint points show every condition and subset size up to half of its
runs (bias factors 5--20). Dotted lines show typical errors of 1 and 0.5 $\ln$ units. Data for all four CVs are shown in 
Fig.~\ref{fig:s_front} of the supplementary material.}\label{fig:front}
\end{figure}

In Fig.~\ref{fig:cost} we demonstrated that rate constants can be accurately estimated with EATR or EATR-flooding given sufficient simulation expense.
In Fig.~\ref{fig:front}, we show the root-mean-square error in the log of estimated rate sampled over random subsets of data for every individual choice of protocol studied here for the best and worst CVs, with results for all four CVs shown in  Fig.~\ref{fig:s_front}. 
From these data, we can show the Pareto front, which shows the error-cost tradeoff, and from this we can ask whether any guidance could help us reach this front. 

The front is constructed by identifying the best ``strategy'' at each amount of cost. A strategy consists of  a protocol, an estimator and a number
of runs (per condition for WT-MetaD, and per bias level for the flooding protocols).
The error
is the root-mean-square error of $\ln k$ computed over every independent random subset of
that size drawn from every condition of that protocol.
The cost is the median
total simulated time of such a subset, counting production runs only. Fits to a complete
data set are excluded, because a single fit has no spread and the best of several hundred
would be chosen by luck. Subset sizes above half of the runs available are excluded as
well, because subsets drawn from most of a collection of runs share most of their data and understate the
spread of that many independent runs.
The front is the set of strategies that no other strategy can beat on error at the same or lower cost.

First, we can ask which protocols reach this minimal typical level of error at a given cost. 
For $Q$, naive estimators hold the front up to 3.2~$\mu$s: iMetaD with 5--30 runs per
condition and naive OPES-flooding with 5--15 runs per level, as the RMSE falls from 0.90 to
0.20. 
As discussed in earlier sections, EATR fitted to the same runs has a higher error, although within an acceptable range. EATR-flooding only becomes part of the front for VES-flooding at high expense, where the minimum observed error is found.
In contrast, for $R_{ee}$, the front is
WT-MetaD EATR alone, from 5 to 80 runs (0.73--11.6~$\mu$s, RMSE 1.72 to 0.38); no naive strategy has an RMSE below 1.4 for any amount of simulation. 
As is also visually evident in earlier plots, EATR-flooding applied to quasi static bias requires more expense to reach the same level of error, although the extra ability to mitigate failures can lead one to choose these strategies over MetaD.

For the WT-MetaD strategies highlighted on the front, what are highlighted are not the best parameter choices in hindsight but rather the
conditions a practitioner might have chosen in advance given sufficiently good estimates using the bias growth law described in the supplementary
material (Sec.~\ref{sec:s_design}).
This formula takes the width and relaxation time of the CV in the
reactant basin, its apparent barrier, the $\gamma$ of a static-bias pilot (here the five-level OPES-flooding fit at pace 1000) and an estimate of $k_0$ (for
which the reference rate was used in this case).
The growth law then predicts the fill and escape time of every candidate bias
factor, hill height and pace.
As described in the next section, we here can choose a condition where the bias leaves at least
2~$\kB T$ of barrier standing, the pace is at least three tenths of the relaxation time and
the predicted boost is at least 3. 
The transition barrier is the one input that a basin trajectory
does not supply. Here it is extracted from the unbiased profiles of Fig.~\ref{fig:system}; in
practice it would have to come from a free-energy profile along the CV reconstructed by reweighting an
exploratory metadynamics or OPES run.

\subsection{Failure conditions for EATR estimates}\label{sec:works}

\begin{figure}[t]\centering
\includegraphics[width=\columnwidth]{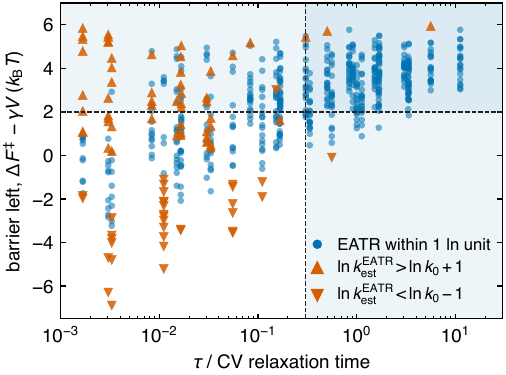}
\caption{Monitoring accuracy of EATR with a growing bias. Each point is one WT-MetaD grid
condition (all four CVs), placed by its deposition interval $\tau$ in units of the CV
relaxation time $\tau_{\rm rel}$ and by the barrier its bias left standing,
$\Delta F^\ddagger-\gamma V_{\rm obs}$, with $\Delta F^\ddagger$ the apparent barrier along
that CV, $V_{\rm obs}$ the deepest fill reached and $\gamma$ the fitted efficiency. Blue
circles show conditions where EATR is within one $\ln$ unit of the reference. Vermilion triangles show those outside it,
pointing up where the estimated rate is too fast,
$\ln k_{\rm est}^{\rm EATR} > \ln k_0 + 1$, and down where it is too slow,
$\ln k_{\rm est}^{\rm EATR} < \ln k_0 - 1$. Nearly all too-slow
failures lie below zero barrier left, where the bias overfilled the basin; nearly all
too-fast failures lie below a tenth of the relaxation time, at any fill. Dashed lines mark
2~$\kB T$ left and $\tau=0.3\,\tau_{\rm rel}$; in the darker region, where both hold, 271
of 273 conditions are accurate.}\label{fig:design}
\end{figure}

Within our data set, we have a small fraction of conditions where it is not possible to obtain accurate kinetics with EATR. 
For protocols with growing bias, we see failures arising via two distinct (and previously known) mechanisms.
If the bias fills the entire well before escape then the process is no longer a thermally activated transition and the recovered rate is too
slow. 
On the other hand, if hills are deposited faster than the CV can relax, the quasi-static Kramers'-like picture is not valid, additional energy is dissipated without increasing the observed rate such that $\gamma$ is reduced, and the estimated EATR-rate constant is too fast.

To show these trends, we show in Figure~\ref{fig:design} a scatter plot of all simulation conditions and highlight those that meet our accuracy threshold of 1 $\ln$ unit.
On the horizontal axis, we show the deposition rate, which we scale by the estimated CV-relaxation time (Sec.~\ref{sec:relax}),
and on the vertical axis, we estimate how much barrier the bias left standing: the apparent barrier along that CV minus $\gamma$ times the deepest fill, which the smallest recorded hill marks, with the fitted $\gamma$ converting fill along the CV into barrier lowering.
In general, rates that are too slow come from overfilled conditions whereas rates that are too fast arise from
deposition rates faster than CV relaxation.
Conditions that retain a few $\kB T$ of barrier and deposit no faster than the CV relaxes have low error almost in all cases.
The dashed lines show heuristic cutoffs that make this quantitative: with less than 1~$\kB T$ left, EATR is accurate in 70\% of conditions,
against 94\% with 2.5--4~$\kB T$; with $\tau$ below 1\% of the relaxation time, in 53\%,
against 99\% for $\tau\ge0.3\,\tau_{\rm rel}$. Requiring at least 2~$\kB T$ left and $\tau\ge0.3\,\tau_{\rm rel}$ retains 273 of 639 conditions, of which 271 are accurate (Sec.~\ref{sec:s_design}). 

We can observe the fast-deposition failure by seeing $\gamma$ change substantially as pace is varied, and so adding a set of data at a single faster pace to one's simulations could provide this check at relatively low cost. 
In standard WT-MetaD (i.e. not variants that cap bias \cite{dama2015exploring}), it is difficult to control over-biasing independent of lowering the pace a priori.
We note that one can derive an approximate growth law for the bias that is a function of bias factor, hill height and pace to 
predict the asymptotic fill level, which can then be used to help decide these parameters, as discussed in Sec.~\ref{sec:s_design}. 

For a static or quasi-static bias, overfilling is the primary failure mode for EATR-flooding. 
While we know that several sets of data with maximum bias levels below the barrier should be used, the true barrier or barrier within a given CV is not always known accurately.
As previously discussed \cite{mazzaferro2026stepping}, 
curvature in $\ln k_{\rm obs}$ versus
$\ln\langle e^{\beta V}\rangle$ can be a signature of overfilling, allowing those sets to be dropped, e.g. as seen for $R_{g}$ in Fig.~\ref{fig:system}d. 
Further information about these results is shown in Figs.~\ref{fig:s_hb}, \ref{fig:s_slopes}, and \ref{fig:s_sets}.

\section{Conclusions}\label{sec:conclusions}

Unbiased kinetics can be recovered reliably from either a slowly growing bias or a static
bias held below the barrier. 
For this benchmark system, both classes of method estimate rate constants with accuracy to one $\ln$ unit for a wide range 
of parameter choices within the methods. 
Estimator choice matters more than biasing scheme for all but the best choice of CVs. 
Selecting a poor CV results in part of the bias being wasted, and this must be factored into the estimate.
EATR and EATR-flooding account for this biasing efficiency $\gamma$ from a
growing bias or several static levels. 
The fitted $\gamma$ then separates the regimes: above
$\gamma\approx0.8$, a naive estimator is slightly more precise if the rescaled times are
confirmed exponential; below it, fitting $\gamma$ is essential. 
The same parameter can also be used to diagnose deposition that is too fast, because a coordinate property should not change
with pace.

Because $\gamma$ is a property of the coordinate more than of the protocol (Fig.~\ref{fig:cost}c), we suggest that it can be roughly estimated before a CV is chosen. 
A few short flooding runs at each of two static levels well below the barrier, fitted together with
EATR-flooding, give the slope of $\ln k_{\rm obs}$ against $\ln\langle e^{\beta V}\rangle$
(Fig.~\ref{fig:system}d), and with it $\gamma$ and a first estimate of $k_0$. Repeated on
each candidate coordinate, this screen ranks CVs before any production simulation: a
coordinate with $\gamma$ near one can be biased with any scheme and analyzed with either
estimator, while one with $\gamma$ well below one requires EATR or argues for a better
coordinate. The same $\gamma$ and $k_0$ can be used as inputs to an approximate growth law that can help to choose
well-tempered parameters.

One downside of benchmark systems where unbiased rates can be recovered by exhaustive sampling is that the maximum boost in efficiency from biased rate estimates is low, given that there is only a 6~$\kB T$ barrier to overcome.
Obtaining accurate rate constants for our benchmark system ends up taking about 1--20 mean first-passage times of the unbiased system, which is not much more efficient than performing unbiased simulations.
However, we know that the advantage should grow exponentially with the barrier.
To confirm that for this system, in the SI we report results for studying this system at 270~K, which raises the barrier along $Q$ from 6.0
to 16.1~$\kB T$, which predicts, with no free parameter, a rate ten $\ln$ units slower.
Although we cannot compute the unbiased rate, we can demonstrate that EATR-flooding on $Q$ remains accurate all the way to this low temperature with a mean
offset of $+0.26\pm0.44$ $\ln$ units over the three lower temperatures, while the naive estimate drifts to $+1.6$ at 270~K
(Sec.~\ref{sec:s_temperature}). 
When a single WT-MetaD condition per temperature was instead applied along $R_{ee}$, EATR incurred a larger error (up to 1.9 $\ln$ units), but it is orders of magnitude more accurate than iMetaD, which is low by 4--7 $\ln$ units.
At 270~K the $Q$-biased runs escape on average $7\times10^5$ times faster than the unbiased mean first-passage time, and the five levels used to compute this estimate cost 6~$\mu$s against an expected unbiased unfolding time of 16~ms. 

The overall message conveyed by our data is simple: measure $\gamma$, leave part of the barrier to allow for an activated process, and deposit no faster than the CV relaxes in the reactant basin.
While there are no fixed rules for what to choose for a new system nor can we say precisely how many simulations will be needed to obtain good fits, the barrier- and CV-relaxation-time conditions are directly testable in atomistic systems to which EATR has since been applied.\cite{mazzaferro2026stepping,mazzaferro2025using} 
The performance of EATR as well as optimal choices of protocols for processes with multiple sequential barriers or nonactivated dynamics remain open, where one $\gamma$ per coordinate or a single-rate survival model may not be sufficient.
Which biasing schemes or estimators are viable for those systems remains an intriguing subject for future study.

\section{Computational details}\label{sec:compdetails}

\subsection{Molecular dynamics}
We represented the B1 domain of protein G (PDB 1PGB)\cite{gallagher1994two} with the
C$_\alpha$ G\=o-like model of Karanicolas and
Brooks.\cite{karanicolas2002origins,karanicolas2003improved} Each of the 56 residues is one
C$_\alpha$ bead; harmonic bonds and angles, Fourier dihedrals and native-contact
attractions define the force field. Simulations used
LAMMPS\cite{thompson2022lammps} with PLUMED,\cite{plumed2019promoting} a 10~fs time step and
a Nos\'e--Hoover chain thermostat\cite{martyna1992nose} at 312~K (1~ps damping time; chain length 3).
Each trajectory began from the energy-minimized folded structure with random velocities at
the target temperature (LAMMPS \texttt{velocity create} with a uniform distribution and a
different seed for each run) and ended at the first check, every 100 steps, with $Q<0.35$. Reference runs
used the same settings without bias.

\subsection{Well-tempered MetaD}
The grid combines bias factors $b \in \{2, 5, 10, 15, 20\}$, initial hill heights
$h_0 \in \{0.04, 0.10, 0.19, 0.39\}~\kB T$ (0.1, 0.25, 0.5 and 1.0~kJ/mol in PLUMED
units) and deposition paces
$\tau \in \{100, 500, 10^3, 5\times10^3, 10^4, 3\times10^4, 5\times10^4, 10^5\}$ steps,
that is 1~ps to 1~ns. The hill width is 0.02 in the units of the CV (0.06~nm for
$R_{ee}$). Every condition with bias factor 5--20 has 200 runs (one, $R_g$ at $b = 5$,
$h_0 = 0.04~\kB T$, $\tau = 100$, has 199). At bias factor 2 the two largest hill heights
have 50 runs, as does RMSD at $h_0 = 0.10~\kB T$ (163 at $\tau = 100$); the rest have 200.
Of the 640 conditions, 639 yielded a fit. The remaining condition (RMSD, $b=2$,
$h_0=0.10~\kB T$, $\tau=500$) could not be fit; it was therefore treated as a failure and is not included in the quantitative analysis.

\subsection{OPES-flooding}
OPES-flooding used \texttt{OPES\_METAD} with a kernel width of 0.02 (0.06~nm for
$R_{ee}$), \texttt{BARRIER} $\Delta E \in \{1.9, 2.7, 3.5, 4.2, 5.0\}~\kB T$ (5 to 13~kJ/mol) at the same eight paces, and
50 runs per level. The excluded region suppressed deposition at $Q < 0.65$ when biasing
$Q$ or RMSD, at $R_g > 1.15$~nm when biasing $R_g$ and at $R_{ee} > 2.9$~nm when biasing
$R_{ee}$. Additional levels were run at pace 1000: at 5.8 and 6.6~$\kB T$ for $Q$ and $R_{ee}$, and at 5.8, 6.6, 7.3 and 8.1~$\kB T$ for RMSD and $R_g$.

\subsection{VES cutoff flooding}
VES flooding proceeded in two stages. In stage 1, one optimization run per cutoff
expanded the bias in symlet wavelets (\texttt{BF\_WAVELETS}, order 10, 26
functions)\cite{pampel2022improving} over $[0, 1]$ for $Q$, $[0, 3.6]$~nm for RMSD,
$[0, 3.8]$~nm for $R_g$ and $[0, 10.5]$~nm for $R_{ee}$. The target was uniform with a
Fermi-switched cutoff ($\lambda = 10$), and the coefficients were optimized by averaged
stochastic gradient descent with step 0.1 and update stride 500. In stage 2 the
converged coefficients were applied as a static bias, with
$F_c \in \{1.9, 2.7, 3.5, 4.2, 5.0, 5.8\}~\kB T$ (5 to 15~kJ/mol) and 200 production runs
per cutoff.

\subsection{Relaxation times of the collective variables}\label{sec:relax}
We measured relaxation within the folded basin from the unbiased first-passage runs, which
record $Q$, $R_{ee}$ and $R_g$ every 10~ps. 
For a subset of 38 of these runs, we truncated each at its first excursion
below $Q=0.65$, which left reactant-basin segments of 6--290~ns, discarding segments shorter than 5~ns. 
For each CV we then
averaged the autocorrelation of the fluctuations about the segment mean and integrated it
to its first zero crossing, giving 310~ps ($Q$), 330~ps ($R_{ee}$) and 90~ps ($R_g$). 
By comparison, using all 400 runs, with 374 such segments, gives about 340, 370 and 110~ps.
RMSD was not recorded in those runs, so we used the first 30~ns of 20 grid runs at
the slowest pace and smallest hill height, where the deposited bias has not yet reached
0.6~$\kB T$, which gives 610~ps. The estimates are uncertain by roughly a factor of two:
the spread over segments is 80--230~ps, the same nearly unbiased segments give 480, 360
and 180~ps for $Q$, $R_{ee}$ and $R_g$, and the time at which the autocorrelation falls
to $1/e$ is shorter again. Nothing below depends on more than the order of magnitude.

\subsection{Rate fits}
The cumulative distributions of all MetaD simulations were fitted by binomially weighted least
squares, while the flooding rate calculations were performed with uniformly weighted fits. 
To accelerate the calculations, an option in the EATR-rate-analysis package was used to fit every 1000th row of the output COLVAR files, with a smaller stride where needed so that the shortest trajectory of a data set preserved at least 200 rows (or was not strided if fewer than 200 rows were available).
Runs that did not
transition entered the fits as censored. When using bootstrapping for error analysis, the estimate reported is the mean over bootstrap samples of the trajectories, not the single fit to
all runs. 
On the WT-MetaD grid, the two differ by a median of $-0.08$ ln units for EATR  and by at most 0.15 for iMetaD. Uncertainties are 95\%
percentile intervals of the resamples for the grid, and bootstrap standard deviations for
the flooding fits and the reference. All fits used the EATR-rate-analysis package. 
The per-set quantities come from a fit to all runs of each level without resampling.

\section*{Supplementary Material}
The supplementary material reports condition-level accuracy over the MetaD grid, the
minimum-data analysis, flooding-level fits and overfilling screens, biased landscapes,
VES validation, the growth-law and prospective tests, extended OPES levels, pace and
frequency-adaptation tests, cases in which EATR underperforms iMetaD, and censored runs. Tables of all individual fits are provided in the analysis repository.

\begin{acknowledgments}
This study was primarily funded by the National Institutes of Health under award
number R35GM138312, which provided support for T.P., N.M. and G.M.H.
N.M. was also partially supported by a fellowship from the Simons Center for
Computational Physical Chemistry (SCCPC, Simons Foundation Grant MPS-T-MPS-00839534, MET)
and a Margaret Strauss Kramer Fellowship from the Department of Chemistry at NYU. 
G.M.H. also gratefully acknowledges support from the Alfred P. Sloan foundation through a Sloan Research Fellowship. 
P.C. was supported by the Flatiron Institute, a division of the Simons Foundation.
G.M.H. and N.M. thank the Center for Computational Mathematics at the Flatiron Institute
for their hospitality while a portion of this research was carried out.
Finally, this work was enabled through the NYU IT High Performance Computing resources, services, and staff expertise.
\end{acknowledgments}

\section*{Author Declarations}
\subsection*{Conflict of Interest}
The authors have no conflicts to disclose.

\subsection*{Author Contributions}
G.M.H. and P.C. designed the study. T.P., N.M. and G.M.H. ran the simulations. T.P. and G.M.H. analyzed the data. T.P., G.M.H. and P.C. wrote the paper.

\section*{Data Availability}
The data, simulation inputs, analysis scripts, a notebook that draws every figure, and tables of every individual
fit behind the figures are available at \url{https://github.com/hocky-research-group/best-rate-strategy-data}, which also links to the full simulation output. All fits can be reproduced with the EATR-rate-analysis package
(\url{https://github.com/hocky-research-group/EATR-rate-analysis}).

\bibliography{EfficientEATR}

\clearpage
\onecolumngrid
\raggedbottom
\makeatletter
\let\set@footnotewidth\set@footnotewidth@one
\let\compose@footnotes\compose@footnotes@one
\expandafter\let\csname combine@insert@1\endcsname\combine@insert@@ne
\makeatother
\setcounter{section}{0}\setcounter{figure}{0}\setcounter{table}{0}\setcounter{equation}{0}\setcounter{footnote}{0}
\renewcommand{\thefigure}{S\arabic{figure}}
\renewcommand{\thetable}{S\arabic{table}}
\renewcommand{\theequation}{S\arabic{equation}}
\renewcommand{\theHsection}{SM.\arabic{section}}
\renewcommand{\theHfigure}{SM.\arabic{figure}}
\renewcommand{\theHtable}{SM.\arabic{table}}
\renewcommand{\theHequation}{SM.\arabic{equation}}
\begin{center}
{\large\bfseries Supplementary Material for: Efficient strategies for obtaining accurate kinetics from biased simulations\par}
\vspace{1.5ex}
Triasha Pal, Nicodemo Mazzaferro, Pilar Cossio and Glen M. Hocky
\end{center}
\vspace{1ex}
\section{Prospective accuracy rule for a growing bias}\label{sec:s_design}
Below, we will show that a growing bias is reliable when it leaves part of the barrier intact and deposits slowly
enough for the CV to relax between hills. This section turns those requirements into a
procedure that can be applied before production, then validates every step retrospectively
and prospectively.

\emph{Procedure.}
\begin{enumerate}
\item From a short trajectory confined to the reactant basin, estimate the CV width
$\sigma_A$ and relaxation time $\tau_{\rm rel}$. Obtain an approximate apparent barrier
$\Delta F^\ddagger$ along the same CV, from a free-energy profile reconstructed by
reweighting an exploratory metadynamics run (or via quasi-static biasing as used in step 2).
\item Measure the biasing efficiency $\gamma$ and obtain an initial rate $k_0$ from
static-bias runs at several levels below the barrier, fitted together with EATR-flooding.
Below, $\gamma$ is the EATR-flooding value from the OPES levels at pace 1000, and $k_0$ is
the reference rate.
\item For each candidate bias factor $b$, initial hill height $h_0$ and pace $\tau$, use
Eq.~\eqref{eq:s_growth} to solve for the predicted hill count $n^\ast$, fill
$V_{n^\ast}$, escape time $t^\ast=n^\ast\tau$ and boost $1/(k_0t^\ast)$. Retain only
candidates that satisfy Eq.~\eqref{eq:s_rule}.
\item Run production with EATR and repeat at a second acceptable pace. Accuracy is
supported when the rate and $\gamma$ agree across paces and their bootstrap intervals are
stable as trajectories are added. Report $\gamma$, the predicted barrier left and
$\tau/\tau_{\rm rel}$ with the rate.
\end{enumerate}

\subsection*{Approximate escape time from a MetaD bias and effect on kinetics.} A hill of width $\sigma$ overlaps a basin of width $\sigma_A$, giving
an effective increment $h_{\rm eff}=h_0\sigma/\sqrt{\sigma^2+2\sigma_A^2}$. After $n$
hills, the bias at the deposition point is\footnote{N. Mazzaferro, D. T. Limmer, P. Cossio and G. M. Hocky, manuscript in preparation.}
\begin{equation}
  V_n = (b-1)\kB T\ln\!\left[1 + \frac{n h_{\rm eff}}{(b-1)\kB T}\right],
  \label{eq:s_growth}
\end{equation}
with $(b-1)\kB T = \kB\Delta T$ the tempering energy. Equation~\eqref{eq:s_growth} is the
explicit finite-$n$ form of a known asymptotic result: Dama, Parrinello and Voth \footnote{J. F. Dama, M. Parrinello and G. A. Voth, Phys. Rev. Lett. \textbf{112}, 240602 (2014).} prove
that the level of a well-tempered bias grows logarithmically in the number of hills,
while the hill height decays as $1/n$. Here the overlap fixes the
increment, and the law is used before escape rather than at convergence. Escape becomes
likely when $k_0e^{\gamma V_n/\kB T}n\tau=1$. This condition defines $n^\ast$, the escape
time $t^\ast=n^\ast\tau$ and the boost
$1/(k_0t^\ast)=e^{\gamma V_{n^\ast}/\kB T}$. All inputs are available before production.

\begin{figure}[ht]\centering\includegraphics[width=3.37in]{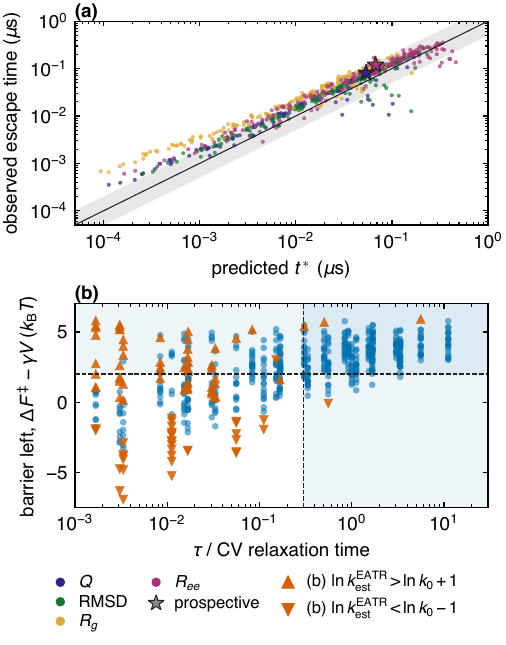}
\caption{(a) Test of the growth law. The observed mean escape time of every grid condition is plotted against
the prediction $t^\ast$ of Eq.~\eqref{eq:s_growth}, using the unbroadened hill overlap and
the fitted $\gamma$; the gray band marks a factor of two. Stars show the two prospective protocols, run
after their predictions were committed, with $t^\ast$ from the OPES-flooding $\gamma$. (b) Every
grid condition is placed by how much barrier its bias actually left standing (fitted $\gamma$,
observed fill) and by its deposition interval in units of the relaxation time of the CV.
Blue circles show conditions where EATR lands within one ln unit and vermilion triangles those where it does
not, pointing up where the estimated rate is too fast,
$\ln k_{\rm est}^{\rm EATR} > \ln k_0 + 1$, and down where it is too slow,
$\ln k_{\rm est}^{\rm EATR} < \ln k_0 - 1$. Light
shading: one of the two accuracy conditions of Eq.~\eqref{eq:s_rule} holds; darker: both hold.}\label{fig:s_design}
\end{figure}

\emph{Accuracy of the prediction (Fig.~\ref{fig:s_design}a).} The smallest recorded hill
marks its deepest fill, $V_{\rm obs} = -(b-1)\kB T\ln[h_{\min}(1-1/b)/h_0]$, since the
well-tempered height falls monotonically with the bias already present. That fill follows
Eq.~\eqref{eq:s_growth} with the unbroadened overlap, the ratio of observed to predicted
bias being 0.90--0.98 for three of the four CVs and 0.62 for $R_g$, whose basin is narrower
than the hill (Fig.~\ref{fig:s_growth}), and the escape times track the prediction over
three decades, 72\% of conditions inside a factor of two (36\% for $R_g$). The untempered
conditions test the same law at $b \to \infty$ (Sec.~\ref{sec:s_growth}).

\emph{Accuracy region (Fig.~\ref{fig:s_design}b).} This panel is retrospective: it places
every grid condition by the barrier its bias actually left,
$\Delta F^\ddagger - \gamma V_{\rm obs}$ with the fitted $\gamma$, and by
$\tau/\tau_{\rm rel}$, with $\tau_{\rm rel}$ the integrated autocorrelation time of the CV
in the basin. For this figure and the counts below, $\Delta F^\ddagger$ is the barrier of the unbiased PMF of each CV (the dashed lines of Fig.~1 of the manuscript), not a reweighted estimate.
With less than 1~$\kB T$ left, EATR is accurate to one ln unit in 70\% of
conditions, versus 94\% with 2.5--4~$\kB T$ left. Below 1\% of the relaxation time, only
53\% are accurate, versus 99\% for $\tau\ge0.3\tau_{\rm rel}$. Requiring both at least
2~$\kB T$ of residual barrier and $\tau\ge0.3\tau_{\rm rel}$ retains 273 of 639 conditions;
271 of those 273 are accurate, versus 76\% outside. The 88 failures outside split into 55
that left less than 2~$\kB T$ of barrier (median error $-1.2$) and 33 that left more but
were deposited too quickly (median $\gamma=0.39$, median error $+1.4$), confirming that both checks are necessary.
Because the four coordinates relax on similar timescales, these data cannot separate the
ratio $\tau/\tau_{\rm rel}$ from a threshold on $\tau$ itself; the relaxation time is the
scale the mechanism implies. The second failure is not announced by the fit: of the 33
failures with barrier left, only two have an EATR Kolmogorov--Smirnov $p < 0.05$. What
does show it is that $\gamma$ tracks the pace, rising by 30--55\% between pace $10^3$ and
$10^5$ for $Q$, RMSD and $R_g$ before it plateaus (Table~\ref{tab:s17}).

\emph{Prospective screen.} Replacing fitted $\gamma$ and observed fill by the OPES-flooding
$\gamma$ and predicted fill gives
\begin{equation}
  \gamma V_{n^\ast} \le \Delta F^\ddagger - 2\kB T, \qquad
  \tau \ge 0.3\,\tau_{\rm rel}, \qquad
  \frac{1}{k_0 t^\ast} \ge 3,
  \label{eq:s_rule}
\end{equation}
The third condition excludes accurate protocols that are too slow to be useful: the four
conditions that pass the physical checks but predict a boost below 3 cost a median
56~$\mu$s, versus 15~$\mu$s for the accepted set. Together, the first and third conditions
bound the predicted boost between 3 and $e^{\beta\Delta F^\ddagger-2}$---50--140 for the
6--7~$\kB T$ barriers here, but $7\times10^7$ for a 20~$\kB T$ barrier.

Using only $\sigma_A$, $\tau_{\rm rel}$, the apparent barrier, $k_0$ and the OPES-flooding
$\gamma$, Eq.~\eqref{eq:s_rule} accepts 57, 48, 70 and 44 of the 128 candidate conditions
for $Q$, RMSD, $R_g$ and $R_{ee}$. All 219 accepted conditions are accurate to one ln unit;
only 62--84\% of rejected conditions are. For $R_g$ and $R_{ee}$, accepted conditions meet
the nine-in-ten subset criterion with 40 runs, at 1.8 and 5.8~$\mu$s;
rejected conditions never meet it. The accepted conditions form the WT-MetaD pool in Fig.~4
of the manuscript.

To test transfer beyond this retrospective grid, two off-grid protocols were registered
before simulation; both meet their predicted accuracy outcome
(Sec.~\ref{sec:s_prospective}).

\emph{Adaptive pace for when the optimal pace is unknown.} Frequency-adaptive
deposition\footnote{Y. Wang, O. Valsson, P. Tiwary, M. Parrinello and K. Lindorff-Larsen, J. Chem. Phys. \textbf{149}, 072309 (2018).} starts fast and lengthens the interval as the acceleration
grows. At matched cost it lands on the fixed-pace curve, neither better nor worse
(Sec.~\ref{sec:s_fametad}), so it buys no accuracy, but it fails more gracefully in its one
parameter than a fixed pace does in $\tau$.

\section{Validity and limits of the growth law from the previous section}\label{sec:s_growth}
\begin{figure}[ht]\centering\includegraphics[width=\textwidth]{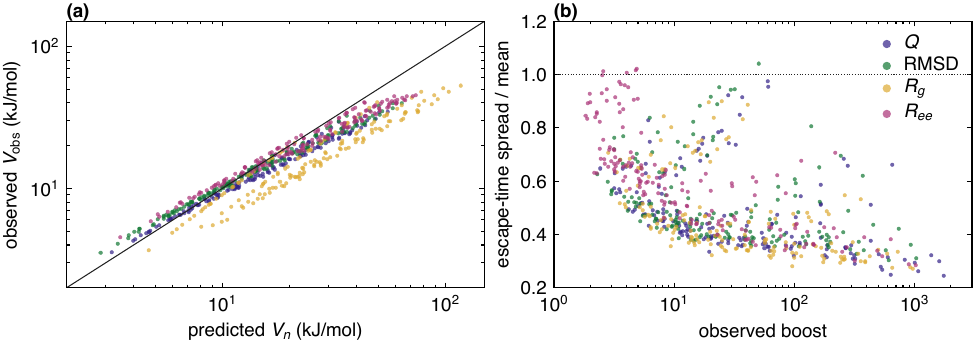}
\caption{(a) The largest bias reached at a deposition point before escape, obtained from the smallest
recorded hill height, is plotted against the growth law with the unbroadened hill overlap.
(b) The spread of the escape times divided by their mean is plotted against the observed boost; the dotted line marks
the value 1 expected for exponential times.}\label{fig:s_growth}
\end{figure}

Figure~\ref{fig:s_growth} tests both the predicted fill and the escape-time distribution.
The observed fill is within 10\% of the prediction for $Q$, RMSD and $R_{ee}$, but 38\%
low for $R_g$, whose basin is narrower than the hill.

Untempered MetaD tests the $b\to\infty$ limit, where the candidate increments make sharply
different predictions. The broadened form,
$h_0\sigma/\sqrt{\sigma^2 + 2 b \sigma_A^2}$, vanishes as $b$ grows and predicts no fill at
all; the unbroadened form used throughout this work,
$h_0\sigma/\sqrt{\sigma^2 + 2 \sigma_A^2}$, does not depend on $b$ and predicts the same
value there as on the grid. On the tempered grid, the fitted increment falls by a factor
1.4--2.6 from $b=2$ to 20, compared with 3.1 predicted by the broadened form. Because
untempered hills do not decay, their fill is read directly from the largest bias before
escape. The linear law $V_n=nh_{\rm eff}$ then gives the increment from fill divided by
hill count.

The unbroadened increment is the accurate one when the CV can relax. Excluding $R_g$, the
observed-to-predicted increment is $1.07\pm0.22$ for the eight conditions with
$\tau\ge0.3\tau_{\rm rel}$ and $0.70\pm0.15$ for the ten faster conditions; it correlates
0.815 with $\ln(\tau/\tau_{\rm rel})$. Thus the pace condition does more than screen rate
estimates: it specifies when the fill model itself is quantitatively valid.

$R_g$ exposes a separate limitation. Its ratio is 0.31--0.57 over the three paces and remains 0.53--0.57 even at
$\tau = 11\,\tau_{\rm rel}$, matching its 38\% shortfall on the tempered grid. Its basin
width, $0.0139$, is smaller than the hill width $0.02$, so the overlap the increment is
derived from is inaccurate. Slower deposition fixes insufficient relaxation, but not a hill
that is too wide for the basin.

\section{How the biased landscape controls accuracy}
\begin{figure}[H]\centering\includegraphics[width=\textwidth]{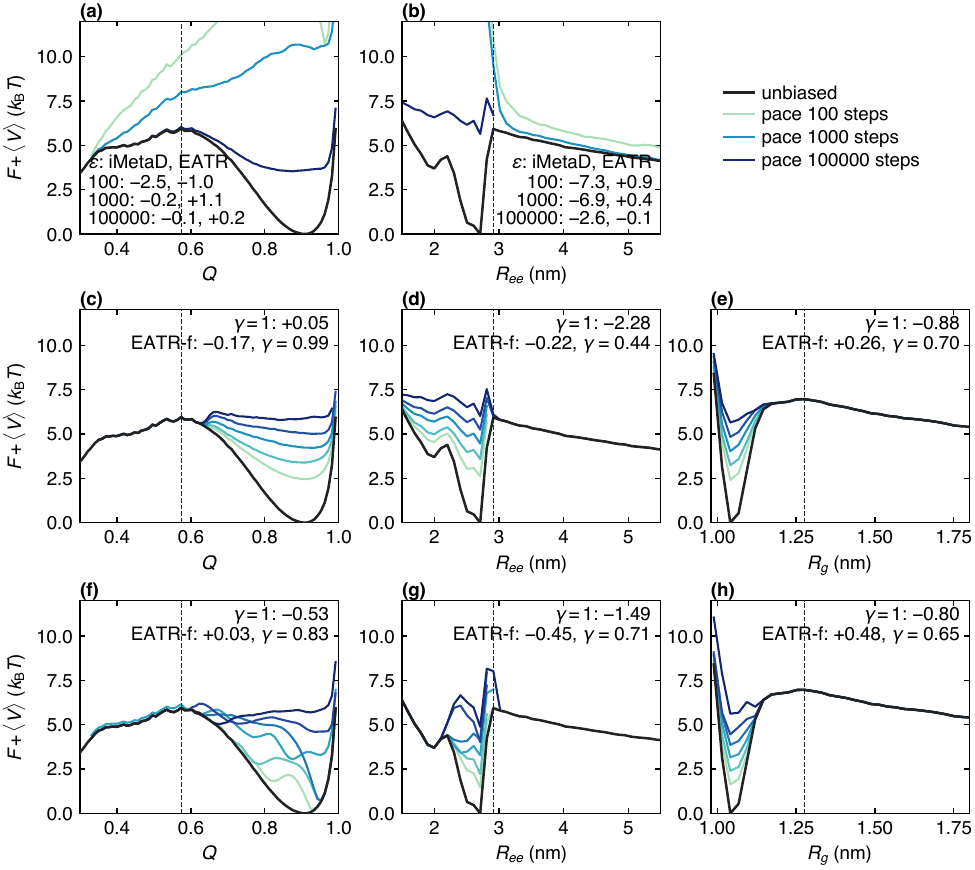}
\caption{The unbiased PMF (black) is compared with $F + \langle V\rangle$ for each bias level, with curves shaded from light to dark for fast to slow pace in (a, b) and for weaker to
stronger bias in (c--h). (a, b) WT-MetaD is shown at $b = 10$ and $h_0 = 0.19~\kB T$ for three paces, with the bias averaged
near escape; the annotations give the errors of iMetaD and EATR on the full condition. (c--e)
OPES-flooding is shown at pace 1000 for \texttt{BARRIER} values of 1.9--5.0~$\kB T$. (f--h) VES cutoff flooding is shown for
$F_c = 1.9$--5.8~$\kB T$. The dashed line marks the apparent barrier.}\label{fig:s_overlays}
\end{figure}

Figure~\ref{fig:s_overlays} compares the unbiased PMF with $F+\langle V\rangle$ for
WT-MetaD, OPES-flooding and VES-flooding. It makes the two main failure modes visible.
Fast WT-MetaD overfills the basin and spills across the barrier. At the slowest pace, the
bias remains below the $Q$ barrier but lifts the $R_{ee}$ basin above its apparent barrier,
consistent with a CV that does not resolve the transition state. 
The static biases (c--h) stay at or below the barrier top for $R_g$ and within 0.3~$\kB T$ of it for $Q$, while on $R_{ee}$ the upper levels lift the flank just inside the barrier to 7--8~$\kB T$, above its 5.9~$\kB T$ top.
$R_g$, biased statically only, sits between the
two, its narrow basin filled to the barrier top without lifting it, with $\gamma$ of
0.65--0.70.

\section{Minimum data and the error--cost front for all four CVs}\label{sec:s_allcv}
\begin{figure}[h]\centering\includegraphics[width=\textwidth]{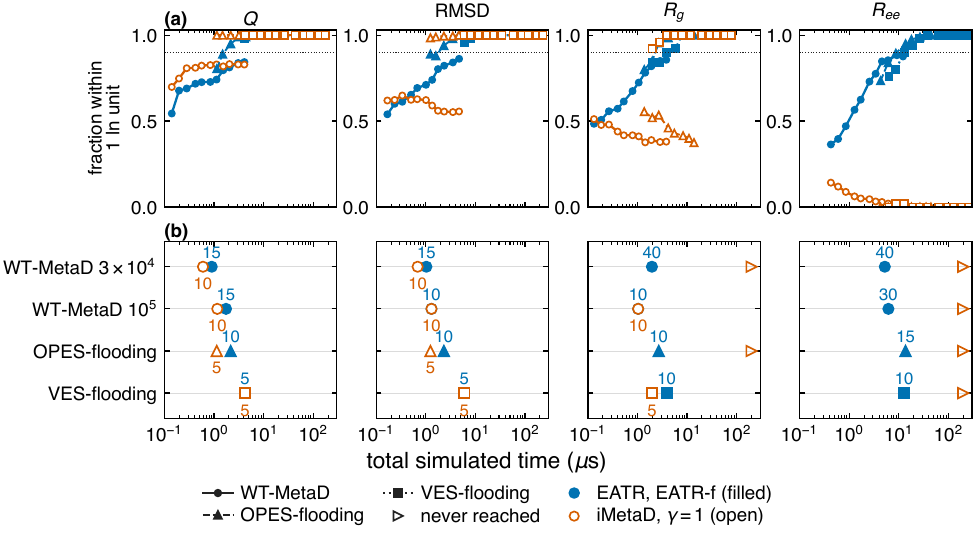}
\caption{Subsampling for every CV. The columns are the four CVs, and the symbols follow
Fig.~3 of the manuscript, where shape and line style give the protocol, and blue with a filled
marker means $\gamma$ was fitted while vermilion with an open marker means $\gamma = 1$. (a)
The fraction of random subsets within one ln unit of the reference is plotted against the median cost of a
subset, and the dotted line marks nine in ten. WT-MetaD is the mean over grid conditions (bias factors 2--20)
at each number of runs, OPES is the mean over paces at each number of runs per \texttt{BARRIER}, and
VES is plotted by runs per cutoff. Subsets are drawn 10 times per grid condition, 25 times per OPES pace
and 50 times for the single VES sweep, more draws where there are fewer conditions to
average over: a CV has 160 grid conditions (159 for RMSD; from 60 runs up only the conditions with more than 50 runs contribute, 144, or 137 for RMSD), eight OPES paces and one VES sweep. (b) The
minimum data is the smallest cost at which that fraction reaches nine in ten and stays there
at every larger size. The number of runs is given beside each symbol (per condition for
WT-MetaD and per bias level for the flooding protocols), and an open arrow at the right edge
marks a case where the criterion is never reached. WT-MetaD is shown at two slow paces,
$3\times10^4$ and $10^5$ steps (300~ps and 1~ns); at $10^3$ steps and faster, neither
estimator reaches the criterion for any CV.}\label{fig:s_mindata}
\end{figure}

\begin{figure}[H]\centering\includegraphics[width=\textwidth]{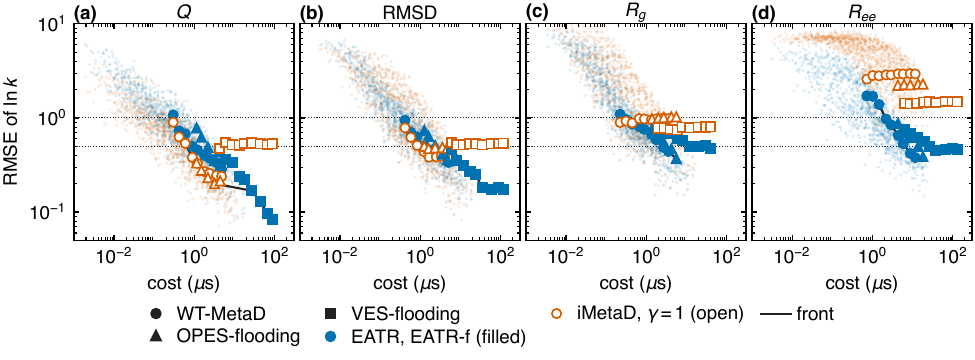}
\caption{Error--cost front by strategy for every CV, as in Fig.~4 of the manuscript. Large
symbols show the typical error of $\ln k$ (root-mean-square over independent subsets) against the
median cost of the production runs, for WT-MetaD on the gentle conditions of the grid
selected in advance by the rule of Sec.~\ref{sec:s_design}, and for the flooding protocols
over all paces. The solid line shows the front. Faint points show every grid condition and subset size up to half of its runs (bias factors 5--20). Three iMetaD points for $R_g$, with errors above 10 ln units, lie beyond the plotted range. Dotted lines show typical errors of 1
and 0.5 ln units.}\label{fig:s_front}
\end{figure}

Figures~3 and 4 of the manuscript show $Q$ and $R_{ee}$, the best and the worst coordinate.
Figures~\ref{fig:s_mindata} and \ref{fig:s_front} add RMSD and $R_g$ and, for the minimum
data, the fraction of subsets within one ln unit from which each minimum is read.

\section{How many subsets are behind each minimum-data entry}
Tables~\ref{tab:s15} and \ref{tab:s15b} give, for every minimum-data entry charted in Fig.~\ref{fig:s_mindata} and for the flooding protocols, how many random subsets were drawn at the chosen size and the uncertainty on the fraction
of them that landed within one ln unit (a 95\% range). With 50 draws per size the VES
entries are resolved to about $\pm 0.1$; the OPES and WT-MetaD entries, with 200
draws, to about $\pm 0.04$.
\begin{table}[H]\centering
\caption{The minimum-data entries of the main text for the growing bias, with their support: cost ($\mu$s) and runs at the smallest size where 90\% of random subsets land within one ln unit and stay there, how many subsets were drawn at that size, and the uncertainty on the fraction of them within one ln unit (a 95\% range).}\label{tab:s15}
\begin{tabular}{lllcccc}\toprule
protocol & estimator & CV & cost ($\mu$s) & runs & draws & 95\% interval\\ \midrule
WT-MetaD, pace $3\times10^4$ & EATR & $Q$ & 0.9 & 15 & 200 & 0.91--0.97 \\
 &  & RMSD & 1.0 & 15 & 200 & 0.86--0.94 \\
 &  & $R_g$ & 2.0 & 40 & 200 & 0.89--0.96 \\
 &  & $R_{ee}$ & 5.2 & 40 & 200 & 0.85--0.93 \\
 & iMetaD & $Q$ & 0.6 & 10 & 200 & 0.92--0.98 \\
 &  & RMSD & 0.7 & 10 & 200 & 0.92--0.98 \\
 &  & $R_g$ & never & -- & -- & -- \\
 &  & $R_{ee}$ & never & -- & -- & -- \\
WT-MetaD, pace $10^5$ & EATR & $Q$ & 1.7 & 15 & 200 & 0.91--0.97 \\
 &  & RMSD & 1.3 & 10 & 200 & 0.88--0.95 \\
 &  & $R_g$ & 1.0 & 10 & 200 & 0.86--0.94 \\
 &  & $R_{ee}$ & 6.2 & 30 & 200 & 0.89--0.96 \\
 & iMetaD & $Q$ & 1.2 & 10 & 200 & 0.88--0.95 \\
 &  & RMSD & 1.3 & 10 & 200 & 0.90--0.97 \\
 &  & $R_g$ & 1.0 & 10 & 200 & 0.89--0.96 \\
 &  & $R_{ee}$ & never & -- & -- & --\\ \bottomrule
\end{tabular}
\end{table}

\begin{table}[H]\centering
\caption{The same for the static bias levels: cost ($\mu$s) and runs at the smallest size where 90\% of random subsets land within one ln unit and stay there, how many subsets were drawn at that size, and the uncertainty on the fraction of them within one ln unit (a 95\% range).}\label{tab:s15b}
\begin{tabular}{lllcccc}\toprule
protocol & estimator & CV & cost ($\mu$s) & runs & draws & 95\% interval\\ \midrule
OPES, 5 barriers & EATR-flooding & $Q$ & 2.2 & 10 & 200 & 0.91--0.97 \\
 &  & RMSD & 2.3 & 10 & 200 & 0.90--0.97 \\
 &  & $R_g$ & 2.7 & 10 & 200 & 0.92--0.98 \\
 &  & $R_{ee}$ & 13.7 & 15 & 200 & 0.91--0.97 \\
 & $\gamma = 1$ & $Q$ & 1.1 & 5 & 200 & 0.98--1.00 \\
 &  & RMSD & 1.2 & 5 & 200 & 0.96--0.99 \\
 &  & $R_g$ & never & -- & -- & -- \\
 &  & $R_{ee}$ & never & -- & -- & -- \\
VES, 6 cutoffs & EATR-flooding & $Q$ & 4.2 & 5 & 50 & 0.90--1.00 \\
 &  & RMSD & 5.9 & 5 & 50 & 0.87--0.99 \\
 &  & $R_g$ & 3.9 & 10 & 50 & 0.79--0.96 \\
 &  & $R_{ee}$ & 12.8 & 10 & 50 & 0.79--0.96 \\
 & $\gamma = 1$ & $Q$ & 4.2 & 5 & 50 & 0.93--1.00 \\
 &  & RMSD & 5.9 & 5 & 50 & 0.93--1.00 \\
 &  & $R_g$ & 2.0 & 5 & 50 & 0.81--0.97 \\
 &  & $R_{ee}$ & never & -- & -- & --\\ \bottomrule
\end{tabular}
\end{table}

\section{Accuracy assessed across all conditions}\label{si:grid}
\begin{figure}[H]\centering\includegraphics[width=\textwidth]{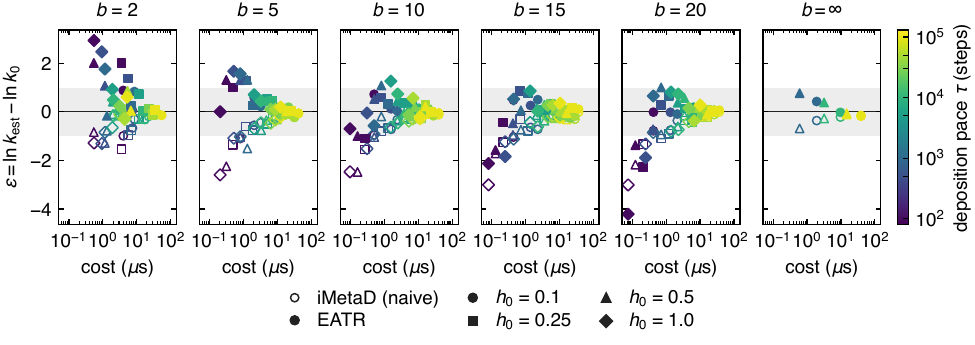}
\caption{Grid conditions for $Q$. The signed error $\varepsilon = \ln k_{\rm est} - \ln k_0$ is plotted
against the cost of the condition (its full data set), with one panel per bias factor $b$; the
$b = \infty$ panel is untempered MetaD, which was run at two hill heights and
three paces rather than the full grid. Open symbols show iMetaD (naive) and filled symbols show
EATR. Shape gives the initial hill height $h_0$ and color the deposition pace $\tau$. The gray band shows one ln
unit.}\label{fig:s_grid_Q}
\end{figure}

\begin{figure}[H]\centering\includegraphics[width=\textwidth]{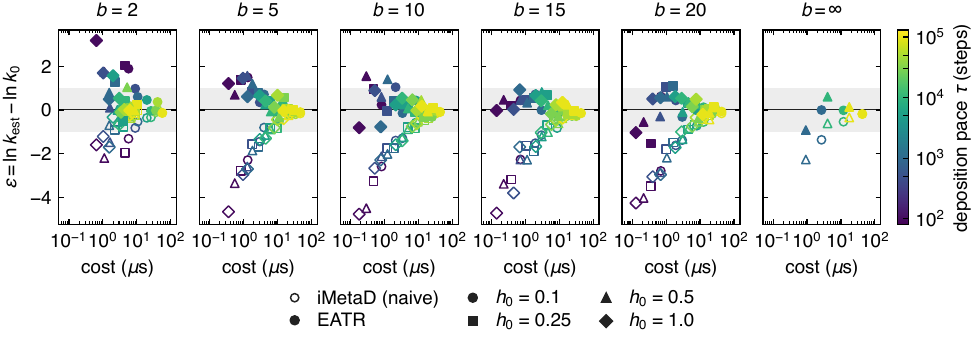}
\caption{Grid conditions for RMSD, as in Fig.~\ref{fig:s_grid_Q}.}\label{fig:s_grid_RMSD}
\end{figure}

\begin{figure}[H]\centering\includegraphics[width=\textwidth]{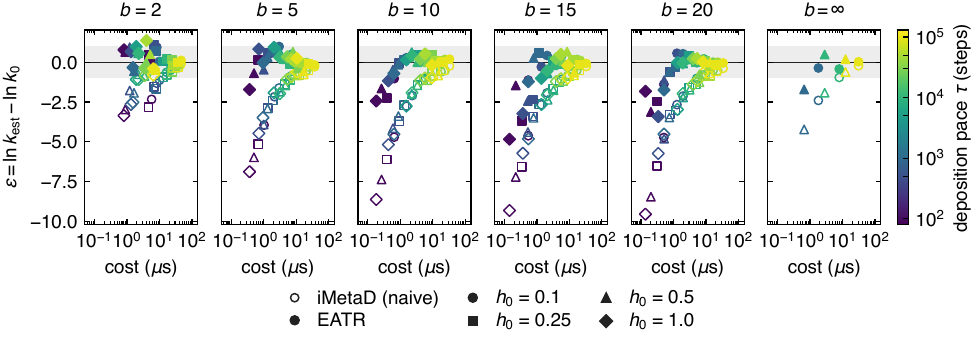}
\caption{Grid conditions for $R_g$, as in Fig.~\ref{fig:s_grid_Q}.}\label{fig:s_grid_Rg}
\end{figure}

\begin{figure}[H]\centering\includegraphics[width=\textwidth]{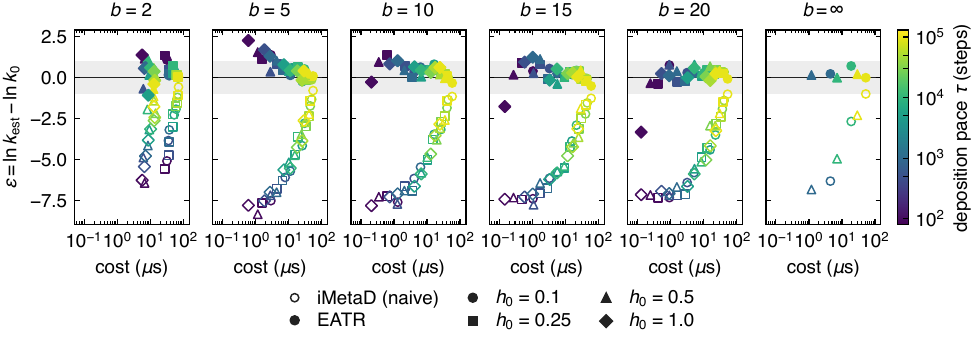}
\caption{Grid conditions for $R_{ee}$, as in Fig.~\ref{fig:s_grid_Q}. The naive estimate is low by up to eight ln units.}\label{fig:s_grid_Ree}
\end{figure}

Figures~\ref{fig:s_grid_Q}--\ref{fig:s_grid_Ree} compare both estimators on every MetaD
condition. Naive error grows as the pace shortens and the CV worsens, reaching eight ln
units for $R_{ee}$ and nine for $R_g$, but depends more weakly on hill height and hardly at all on bias factor. EATR removes most of
this systematic CV dependence. Its remaining scatter is concentrated at the shortest paces
and largest hills, where escape becomes deposition-limited. Pace also sets nearly all of
the computational cost.


\section{Accuracy with censored runs}
\begin{figure}[H]\centering\includegraphics[width=3.37in]{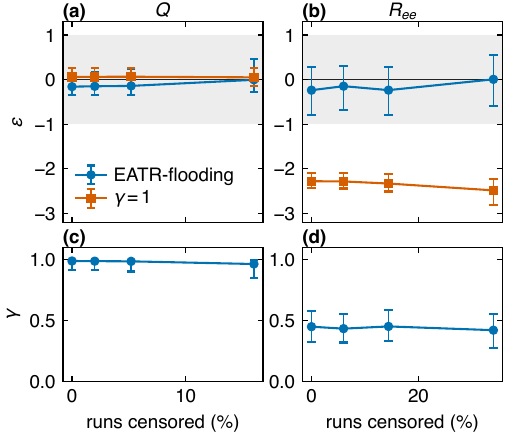}
\caption{Error of EATR-flooding and of the naive ($\gamma = 1$) estimate (top) and the fitted
$\gamma$ (bottom) against the fraction of runs censored. Bars show bootstrap 95\% intervals, and
the gray band shows one ln unit.}\label{fig:s_cens}
\end{figure}

To test whether incomplete trajectories bias the result, Fig.~\ref{fig:s_cens} truncates
the $Q$ and $R_{ee}$ OPES runs at common wall times. The cutoffs censor 10, 25 or 50\% of
the slowest level, corresponding to 2--16\% of all $Q$ runs and 6--34\% of all $R_{ee}$
runs. Treating truncated runs as survivals keeps EATR-flooding inside the uncensored 95\%
interval and moves $\gamma$ by at most 0.03. The naive estimate retains its systematic
bias.

\section{Accuracy across the WT-MetaD conditions}
Tables~\ref{tab:s17} and \ref{tab:s2} report $\gamma$ and full-data accuracy across pace
and bias factor; Sec.~\ref{si:grid} shows every condition. Figure~3 of the manuscript then
asks whether accuracy persists under subsampling: nine of ten random subsets must remain
within one ln unit at every larger sample size. Table~\ref{tab:s4} separates this criterion
by bias factor. Gentle $b=2$ protocols let iMetaD converge with at most ten runs for $Q$, RMSD and $R_g$; for $b=5$--20,
iMetaD still needs only ten runs for $Q$ and RMSD but never meets the criterion for $R_g$. For $R_{ee}$ it never does at any bias factor.
\begin{table}[H]\centering
\caption{Median fitted EATR efficiency $\gamma$ over the well-tempered grid (bias factors 5--20), by deposition pace and, for the paces of $10^4$ steps and slower, by bias factor. The coordinate sets $\gamma$; the bias factor barely moves it, and fast deposition lowers it for every CV but $R_{ee}$, whose $\gamma$ is already the lowest.}\label{tab:s17}
\begin{tabular}{lcccc}\toprule
& $Q$ & RMSD & $R_g$ & $R_{ee}$\\ \midrule
pace 100 & 0.76 & 0.53 & 0.57 & 0.40 \\
pace 500 & 0.73 & 0.57 & 0.60 & 0.38 \\
pace 1000 & 0.71 & 0.58 & 0.59 & 0.37 \\
pace 5000 & 0.80 & 0.69 & 0.66 & 0.36 \\
pace 10000 & 0.82 & 0.76 & 0.72 & 0.34 \\
pace 30000 & 0.89 & 0.88 & 0.73 & 0.33 \\
pace 50000 & 0.91 & 0.86 & 0.74 & 0.36 \\
pace 100000 & 0.93 & 0.89 & 0.81 & 0.39 \\
\midrule $b = 5$ & 0.90 & 0.85 & 0.77 & 0.32 \\
$b = 10$ & 0.90 & 0.83 & 0.73 & 0.34 \\
$b = 15$ & 0.89 & 0.85 & 0.75 & 0.36 \\
$b = 20$ & 0.92 & 0.86 & 0.76 & 0.35\\ \bottomrule
\end{tabular}
\end{table}

\begin{table}[H]\centering
\caption{Median signed error by deposition pace, bias factors 5--20 (iMetaD / EATR for each CV).}\label{tab:s2}
\begin{tabular}{lcccccccc}\toprule
pace (steps) & \multicolumn{2}{c}{$Q$} & \multicolumn{2}{c}{RMSD} & \multicolumn{2}{c}{$R_g$} & \multicolumn{2}{c}{$R_{ee}$}\\ \midrule
100 & -1.63 & -0.57 & -3.37 & 0.45 & -6.54 & -1.68 & -7.42 & 0.81 \\
500 & -0.87 & 0.28 & -1.97 & 0.55 & -3.81 & -0.47 & -7.13 & 0.68 \\
1000 & -0.77 & 0.51 & -1.81 & 0.58 & -2.97 & 0.19 & -6.84 & 0.40 \\
5000 & -0.36 & 0.37 & -1.02 & 0.29 & -1.76 & 0.19 & -5.36 & 0.21 \\
10000 & -0.29 & 0.22 & -0.84 & 0.23 & -1.43 & -0.01 & -4.52 & 0.44 \\
30000 & -0.21 & 0.07 & -0.45 & -0.07 & -0.85 & 0.08 & -3.06 & 0.33 \\
50000 & -0.19 & 0.08 & -0.27 & 0.06 & -0.79 & 0.12 & -2.62 & 0.13 \\
100000 & -0.11 & 0.04 & -0.19 & -0.04 & -0.60 & -0.09 & -1.78 & 0.07\\ \bottomrule
\end{tabular}
\end{table}

\begin{table}[H]\centering
\caption{The same criterion at pace $\ge 10^4$ steps, separately for bias factor 2 and for bias factors 5--20: the gentle bias-factor-2 protocols are where iMetaD converges cheaply.}\label{tab:s4}
\begin{tabular}{llcc}\toprule
bias factor & CV & EATR & iMetaD\\ \midrule
2 & $Q$ & 1.6 (15) & 1.1 (10) \\
2 & RMSD & 2.0 (15) & 1.3 (10) \\
2 & $R_g$ & 3.5 (30) & 0.8 (7) \\
2 & $R_{ee}$ & 19.7 (60) & -- \\
5--20 & $Q$ & 0.8 (15) & 0.6 (10) \\
5--20 & RMSD & 1.0 (15) & 0.6 (10) \\
5--20 & $R_g$ & 1.9 (40) & -- \\
5--20 & $R_{ee}$ & 5.0 (40) & --\\ \bottomrule
\end{tabular}
\end{table}

\begin{figure}[H]\centering\includegraphics[width=3.37in]{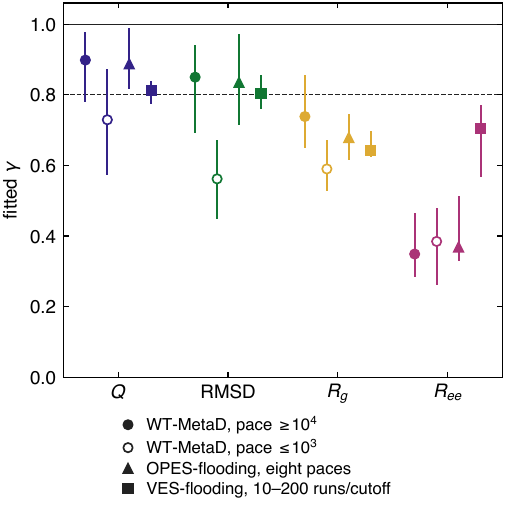}
\caption{The range of the fitted biasing efficiency $\gamma$ for each CV. The median and
10th--90th percentile are shown for WT-MetaD at paces of $10^4$ steps and slower (filled circles)
and of $10^3$ steps and faster (open circles), for the eight OPES paces, and for VES fitted
with 10--200 runs per cutoff. Color is the CV; the pace dependence of the same
fits is shown in Fig.~2c of the manuscript. The dashed line marks $\gamma = 0.8$, above which the naive
estimators are generally accurate.}\label{fig:s_gamma}
\end{figure}

\section{Accuracy with adaptive deposition frequency}\label{sec:s_fametad}
Frequency-adaptive MetaD [Wang, Valsson, Tiwary, Parrinello and
Lindorff-Larsen, \emph{J. Chem. Phys.} \textbf{149}, 072309 (2018)] lets the deposition interval grow with the
acceleration the run has already reached,
\begin{equation}
  \tau_{\rm dep}(t) = \min\left[\tau_0 \max\left(\frac{\alpha(t)}{\theta},\, 1\right),\, \tau_c\right],
  \label{eq:fametad}
\end{equation}
so that the bias fills the basin quickly and then slows before it can reach the transition
state. We therefore compare it with the full fixed-pace error--cost curve, not with one
arbitrarily chosen pace.

The controlled comparison uses $\tau_0=100$ and $\tau_c=10^5$ steps, $b=10$ and
$h_0=0.19~\kB T$. The fixed grid already contains both endpoint paces, so six thresholds
$\theta=0.03$--10 determine where the adaptive runs fall between them. Each threshold has
200 runs on every CV.

\begin{figure}[H]\centering\includegraphics[width=\textwidth]{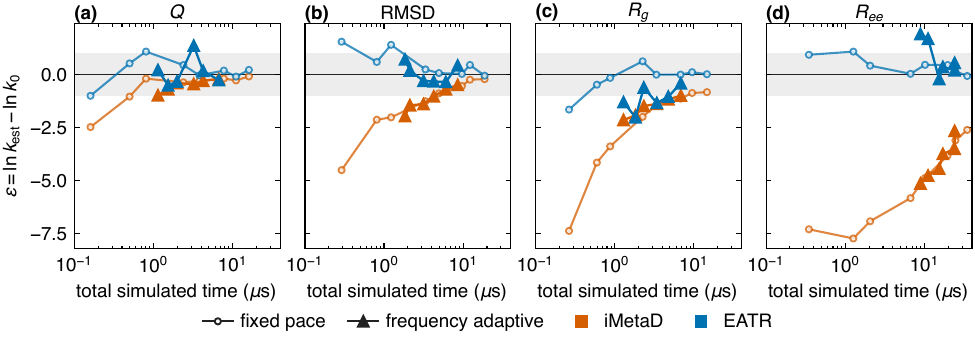}
\caption{Frequency-adaptive deposition against a fixed pace, at bias factor 10 and height
0.19~$\kB T$. Open circles joined by a line show the eight fixed paces of the grid. Filled
triangles show six acceleration thresholds $\theta$ from $0.03$ to $10$ at $\tau_0 = 100$ and
$\tau_c = 10^5$, which are the first and last of those paces. iMetaD is shown in vermilion and EATR in blue. The gray
band shows one ln unit.}\label{fig:s_fametad}
\end{figure}

Adaptive conditions lie on the fixed-pace curve (Fig.~\ref{fig:s_fametad}). At matched
log-cost, their absolute error is $0.17\pm0.08$ ln units higher over 48 comparisons.
Adaptive deposition therefore moves a run along the error--cost curve rather than improving
the curve.

The adaptive family degrades more gently with its aggressiveness parameter, but its most
aggressive setting costs five to seven times more than pace 100, the fastest fixed pace of the grid, for $Q$, RMSD and $R_g$, and 26 times more for $R_{ee}$. Once $\alpha$ exceeds
$\theta$, the interval grows, producing a cost floor of 1.1~$\mu$s versus 0.16~$\mu$s for
fixed pace (the $Q$ values). Equal-cost comparison is therefore the relevant one.

Its advantage is protection against a badly chosen setting. With EATR the worst KS
$p$ value over the conditions of a CV is higher for the adaptive family on three CVs of
four: $0.06$ against $0.001$ for $Q$, $0.285$ against $0.037$ for RMSD and $0.412$ against
$0.155$ for $R_{ee}$, with $R_g$ the same to within $0.001$. A schedule that slows itself
down cannot be pushed as far into the regime where escape times stop being exponential,
which supports its use when a safe fixed pace cannot be chosen in advance.

\section{Prospective test of the design rule}\label{sec:s_prospective}
Two protocols outside the grid in bias factor, height and pace were chosen with the
growth law from $\sigma_A$, the apparent barrier, $k_0$ and the OPES-flooding $\gamma$, by requiring a predicted boost between 3 and 30 and a predicted fill below the barrier, and
their boost, escape time and estimator outcomes were committed to the repository before
any run; their LAMMPS velocity seeds differ from those of the grid. Table~\ref{tab:s13}
compares prediction and outcome. Both protocols also satisfy the three conditions of Eq.~\eqref{eq:s_rule}, which were formulated after these runs, with
margin (a predicted 3.5 and 3.7~$\kB T$ of barrier left, paces of 2.3 and 2.1 relaxation
times, predicted boosts of 12 and 9.6), both
mean escape times are within 5\% of the grid-calibrated expectation, EATR is within one ln
unit for both CVs, and iMetaD is within one ln unit for $Q$ and outside it for $R_{ee}$,
all as registered. The growth-law boost itself is high by a factor of 1.5--1.8, the same
factor by which $t^\ast$ underestimates escape times on the grid.
\begin{table}[H]\centering
\caption{Prospective test of the design rule: two protocols off the grid in bias factor, height and pace, chosen and predicted before any run (predictions committed to the repository first). The expected escape time is $t^\ast$ times the median observed-to-predicted ratio over the grid conditions of the CV that the selection applied to the two protocols accepted, as computed at registration (1.51 for $Q$, 1.86 for $R_{ee}$); errors and $\gamma$ carry bootstrap 95\% intervals, and the OPES-flooding $\gamma$ given to the rule is in parentheses.}\label{tab:s13}
\resizebox{\textwidth}{!}{\begin{tabular}{lcc}\toprule
& $Q$ & $R_{ee}$\\ \midrule
$b$ / $h_0$ (kJ/mol) / pace (steps) & 12 / 0.35 / 70000 & 8 / 0.75 / 70000 \\
runs escaped & 20 of 20 & 40 of 40 \\
boost: predicted / observed & 12.1 / 8.0 & 9.6 / 5.4 \\
escape (ns): $t^\ast$ / expected / observed & 54 / 81 / 81 & 68 / 126 / 120 \\
EATR $\varepsilon$ [95\% interval] & 0.17 [-0.09, 0.62] & 0.58 [-0.22, 1.17] \\
EATR $\gamma$ [95\% interval] (rule input) & 1.00 [1.00, 1.00] (0.99) & 0.30 [0.23, 0.40] (0.44) \\
iMetaD $\varepsilon$ [95\% interval] & 0.20 [-0.06, 0.52] & -2.12 [-4.20, -1.11] \\
EATR within one ln unit (registered expectation held) & yes (yes) & yes (yes) \\
iMetaD within one ln unit (registered expectation held) & yes (yes) & no (yes) \\
KS $p$: EATR / iMetaD & 0.16 / 0.13 & 0.43 / 0.00023\\ \bottomrule
\end{tabular}}
\end{table}

\section{Does a failing Kolmogorov--Smirnov test flag the inaccurate estimates?}\label{sec:s_ks}
\begin{figure}[ht]\centering\includegraphics[width=\textwidth]{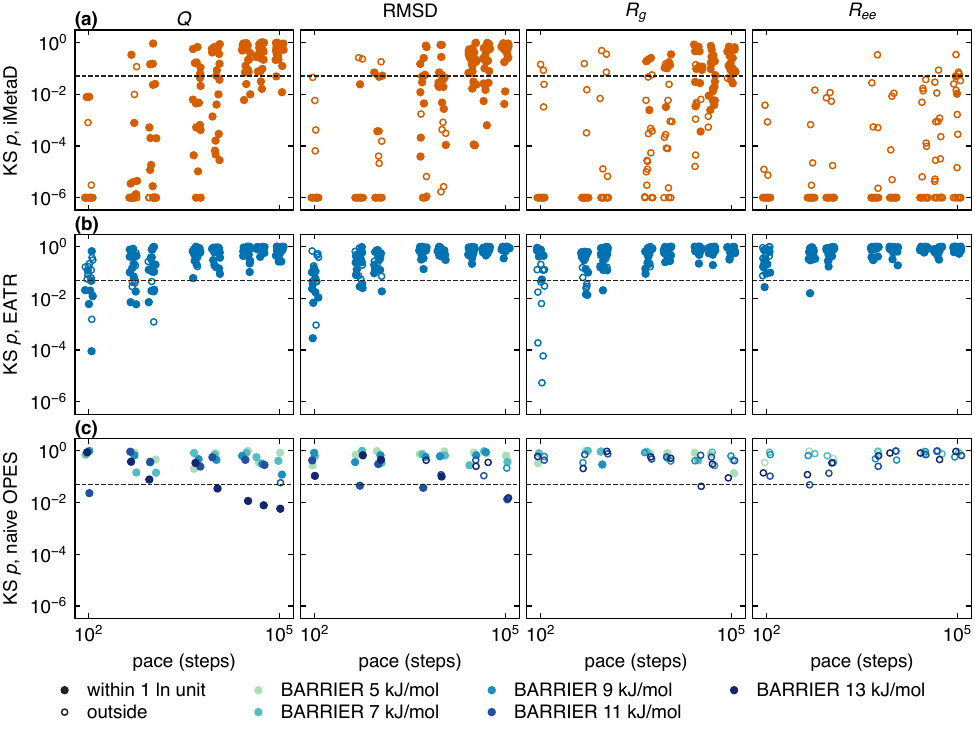}
\caption{KS $p$-value of every full-data fit against deposition pace. Panels (a) and (b) show iMetaD and
EATR on the WT-MetaD grid for bias factors 2--20 and all hill heights, and (c) shows the naive estimate of
each OPES level, coloured by \texttt{BARRIER}. Filled symbols are within one ln unit of the reference, and
open symbols have an unsigned error larger than one ln unit. The dashed line marks $p = 0.05$. Values below $10^{-6}$ are drawn at $10^{-6}$; points are
jittered for clarity.}\label{fig:s_ks}
\end{figure}

Every fit reports a KS $p$-value of its rescaled escape times against an exponential.
Figure~\ref{fig:s_ks} asks whether that test would have identified the estimates that are
off by more than one ln unit, without the reference rate. For iMetaD on the WT-MetaD grid, many of the accurate rate estimates fail the KS test and some inaccurate cases pass.

EATR fails the
test only at paces up to 1000 steps, 24 of its 90 inaccurate fits and 28 of its 549
accurate ones; at paces of 5000 steps and above no EATR fit fails, including the 7
inaccurate ones. The test thus responds to fast deposition and not to overfilling: none of the 54 conditions at these paces with less than 2~$\kB T$ of barrier left fails it (Sec.~\ref{sec:s_design}). 

For a static bias the naive error is the constant shift $(1-\gamma)\ln\langle e^{\beta V}\rangle$,
which leaves the rescaled escape times exponential: of the 74 OPES levels whose naive
estimate is off by more than one ln unit, three fail the test, and the 11 OPES levels that
fail at all sit at \texttt{BARRIER} 11 or 13~kJ/mol, where the bias approaches the
barrier. The VES cutoff ladders behave the same way (not shown): 1 of 8 inaccurate naive
levels fails, and all 4 failures are at the two highest cutoffs.

\begin{figure}[H]\centering\includegraphics[width=\textwidth]{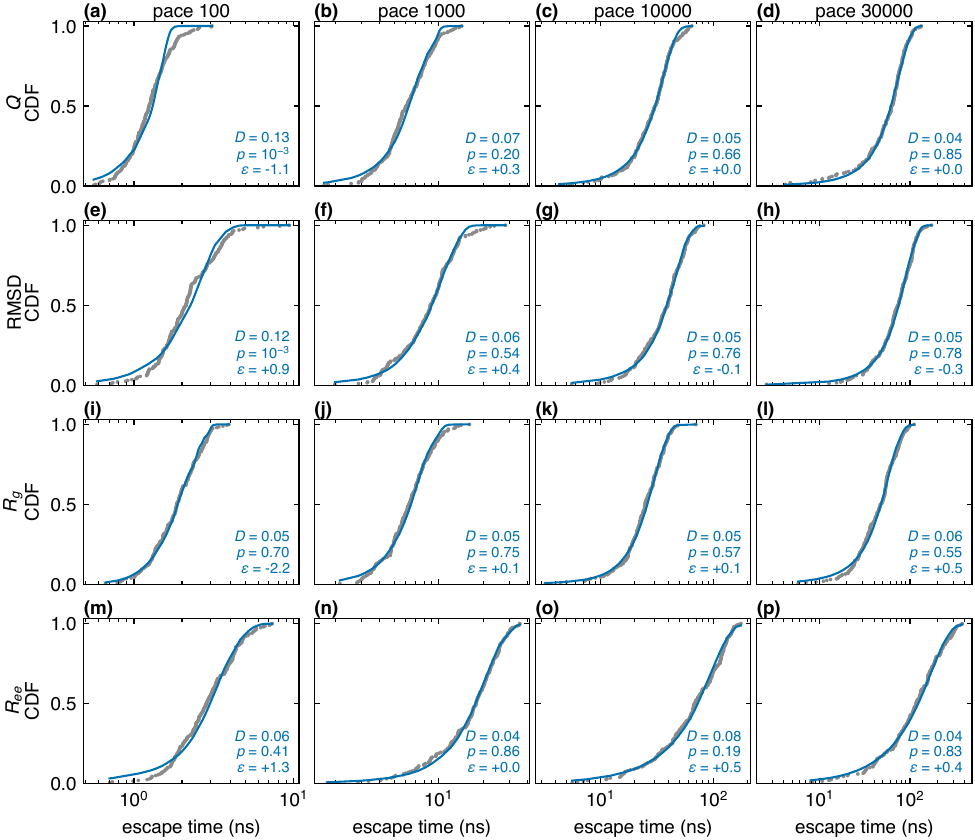}
\caption{Empirical CDF of the escape times (grey) and the EATR fit (blue) against the
escape time, for bias factor 10 and $h_0 = 0.10~\kB T$ at four paces (columns) and every
CV (rows). The annotations give the KS statistic $D$ and $p$-value, and the signed error $\varepsilon$.}\label{fig:s_cdf}
\end{figure}

\begin{figure}[H]\centering\includegraphics[width=\textwidth]{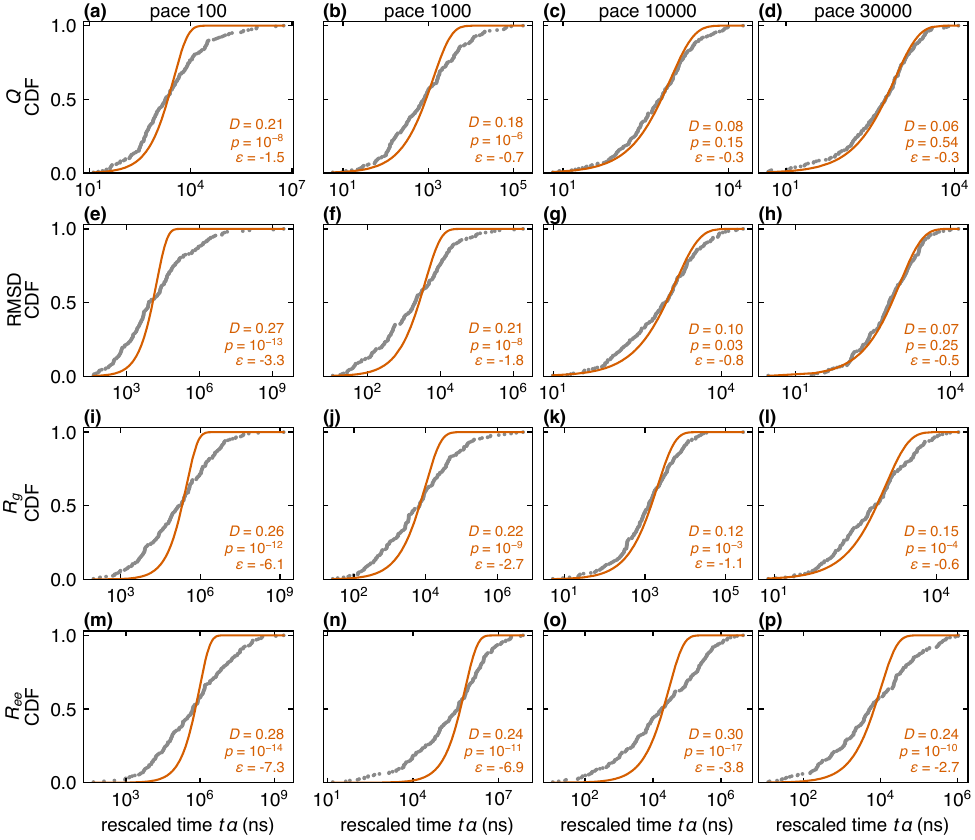}
\caption{The same as Fig.~\ref{fig:s_cdf} for iMetaD, with the empirical CDF (grey) plotted against the rescaled
time $t\,\alpha$ together with the fitted exponential (orange). Note the wider time axes.}\label{fig:s_cdf_imetad}
\end{figure}

Figures~\ref{fig:s_cdf} and \ref{fig:s_cdf_imetad} show examples of the actual curves and fits tested.
For one mid-grid condition (bias factor 10, $h_0 = 0.10~\kB T$) at four paces, the empirical CDF
of the 200 escape times is drawn with the fitted CDF of each estimator: EATR against the
raw escape time, since its fit depends on the raw time through the acceleration of each
run, and iMetaD against the rescaled time $t\,\alpha$, in which its fit is the exponential
$1 - e^{-kt}$. At fast pace the rescaled iMetaD times are broader than an exponential,
with too many early escapes and too long a tail, and that is the failure the test
reports. The EATR fit follows the empirical CDF at every pace. At pace 100 it still fails
the test for $Q$ and RMSD, but passes for $R_g$ and $R_{ee}$ with errors of $-2.2$ and
$+1.3$ ln units: a second fitted parameter can make the escape times exponential without
making the rate right.

\section{Accuracy across static bias levels}
\begin{figure}[H]\centering\includegraphics[width=\textwidth]{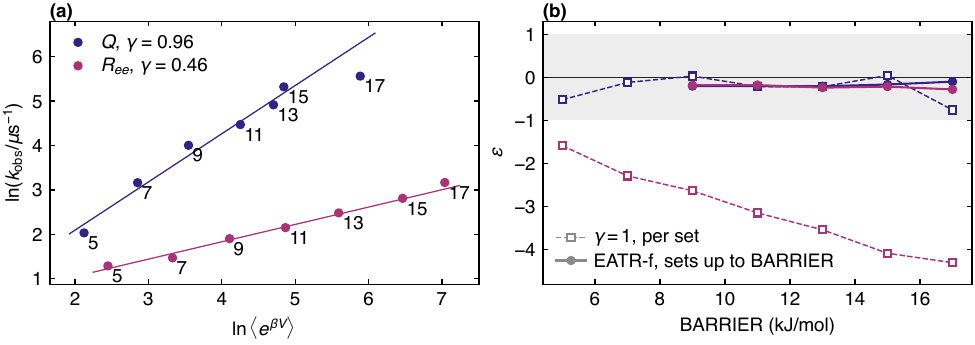}
\caption{OPES bias levels at pace 1000 extended to \texttt{BARRIER} 6.6~$\kB T$ (the apparent barrier along
$Q$ is 5.95~$\kB T$). (a) The observed rate is plotted against acceleration; labels give the \texttt{BARRIER} value, and lines are fits to
the 1.9--5.0~$\kB T$ levels. (b) The error of the naive ($\gamma = 1$) rate estimate of each level (open squares) is compared with that of
EATR-flooding over the levels up to that \texttt{BARRIER} (circles).}\label{fig:s_hb}
\end{figure}

Table~\ref{tab:s5} reports fits using all OPES and VES levels; Table~\ref{tab:s6} applies
the overfilling screen. Figure~\ref{fig:s_hb} then extends the $Q$ and $R_{ee}$ OPES
ladders beyond the apparent barrier to test when the screen succeeds and when fitting
$\gamma$ remains essential.
\begin{table}[H]\centering
\caption{Flooding with every bias level: median (and range) of the signed error and the fitted $\gamma$ for EATR-flooding, and the same error for the $\gamma = 1$ estimate, over the eight paces for OPES (every run) and over the six numbers of runs per cutoff $N$ for VES (10 to 200).}\label{tab:s5}
\begin{tabular}{llccccc}\toprule
protocol & CV & EATRf med $\varepsilon$ & range & $\gamma$ & $\gamma=1$ med $\varepsilon$ & range\\ \midrule
OPES (per pace) & $Q$ & 0.05 & -0.17 / 0.33 & 0.89 & -0.17 & -0.31 / 0.06 \\
OPES (per pace) & RMSD & -0.05 & -0.40 / 0.22 & 0.83 & -0.45 & -0.77 / 0.02 \\
OPES (per pace) & $R_g$ & 0.21 & -0.34 / 0.32 & 0.68 & -1.06 & -1.19 / -0.84 \\
OPES (per pace) & $R_{ee}$ & 0.01 & -0.64 / 0.17 & 0.37 & -2.29 & -2.51 / -2.13 \\
VES (per $N$) & $Q$ & 0.06 & -0.05 / 0.21 & 0.81 & -0.51 & -0.55 / -0.49 \\
VES (per $N$) & RMSD & 0.01 & -0.16 / 0.16 & 0.80 & -0.54 & -0.55 / -0.45 \\
VES (per $N$) & $R_g$ & 0.49 & 0.18 / 0.58 & 0.64 & -0.81 & -0.89 / -0.80 \\
VES (per $N$) & $R_{ee}$ & -0.43 & -0.75 / 0.26 & 0.71 & -1.48 & -1.56 / -1.44\\ \bottomrule
\end{tabular}
\end{table}

\begin{table}[H]\centering
\caption{Overbiasing screen on the flooding fits of Table~\ref{tab:s5} (eight OPES paces; six VES run counts $N$) and on the two ladders extended to 17~kJ/mol: number of fits, how often the screen removed a set, the slope of $\ln k_{\rm obs}$ against $\ln\langle e^{\beta V}\rangle$, and the median absolute error with all sets, with the screened sets, and with the convergence-selected sets.}\label{tab:s6}
\begin{tabular}{llcccccc}\toprule
protocol & CV & fits & screened & slope & $|\varepsilon|$ all & $|\varepsilon|$ screened & $|\varepsilon|$ conv.\\ \midrule
OPES & $Q$ & 8 & 0 & 0.87 & 0.10 & 0.11 & 0.20 \\
OPES & RMSD & 8 & 2 & 0.80 & 0.18 & 0.24 & 0.33 \\
OPES & $R_g$ & 8 & 4 & 0.64 & 0.24 & 0.26 & 0.29 \\
OPES & $R_{ee}$ & 8 & 1 & 0.31 & 0.16 & 0.19 & 0.19 \\
VES & $Q$ & 6 & 1 & 0.81 & 0.07 & 0.08 & 0.11 \\
VES & RMSD & 6 & 0 & 0.81 & 0.08 & 0.08 & 0.32 \\
VES & $R_g$ & 6 & 2 & 0.62 & 0.49 & 0.47 & 0.44 \\
VES & $R_{ee}$ & 6 & 0 & 0.49 & 0.43 & 0.44 & 0.41 \\
OPES 5--17 & $Q$ & 1 & 1 & 0.95 & 0.06 & 0.17 & 0.20 \\
OPES 5--17 & $R_{ee}$ & 1 & 0 & 0.41 & 0.29 & 0.28 & 0.18\\ \bottomrule
\end{tabular}
\end{table}

\begin{figure}[H]\centering\includegraphics[width=\textwidth]{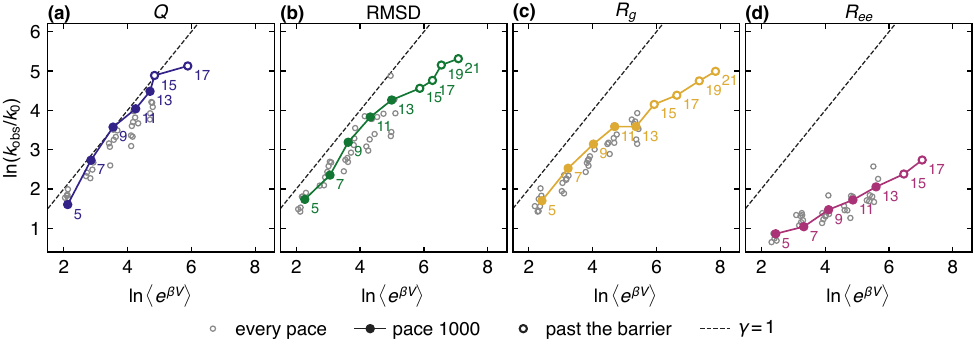}
\caption{Observed rate against acceleration for every OPES level at every pace (grey) and
for the pace-1000 ladder (colour). Open symbols denote levels above 13~kJ/mol, past the apparent
barrier, which were run with the same parameters for these tests; only the $Q$ and $R_{ee}$ levels enter a rate estimate, in Fig.~\ref{fig:s_hb}b and the ``OPES 5--17'' rows of Table~\ref{tab:s6}. Labels give the \texttt{BARRIER} value in kJ/mol. On the dashed diagonal, $\gamma = 1$ and the
naive estimate is exact.}\label{fig:s_slopes}
\end{figure}

\clearpage
Figure~\ref{fig:s_slopes} shows the same construction for all four CVs and every pace.
Below the apparent barrier the five levels fall on a line whose slope is the biasing
efficiency: pace by pace, the fitted slope ranges over 0.73--1.08 for $Q$, 0.66--1.07 for
RMSD, 0.55--0.77 for $R_g$ and 0.27--0.55 for $R_{ee}$, with medians of 0.87, 0.80, 0.64
and 0.31 (Table~\ref{tab:s6}), so a slope
below one, and not curvature, is the ordinary signature of $\gamma < 1$. A downturn
appears only past the barrier, and not on every coordinate: $Q$, whose apparent barrier is 15.4~kJ/mol, falls 0.75 ln units under its line at 17~kJ/mol, while $R_{ee}$, with the same apparent barrier, stays within 0.15 of its line. The RMSD and $R_g$ ladders, whose barriers are
17.9 and 18.1~kJ/mol, were extended to 21~kJ/mol for this test. RMSD bends: its slope
falls from 0.97 below 13~kJ/mol to 0.53 above it, and the 21~kJ/mol level lies 1.1 ln
units under the line through the lower levels. $R_g$ barely does: the slope moves from
0.67 to 0.53 and the top level lies 0.5 under the line, so 3~kJ/mol past the apparent
barrier along $R_g$ the ladder is still close to straight. The Kolmogorov--Smirnov
$p$-values fall below 0.05 only where a level reaches the barrier: for $Q$ and $R_{ee}$ at
15 and 17~kJ/mol (0.01--0.04) and for $R_g$ at 17--21~kJ/mol (0.01--0.02); no RMSD level
is below 0.05.

\section{Stability to the number of fitted bias levels}
\begin{figure}[H]\centering\includegraphics[width=\textwidth]{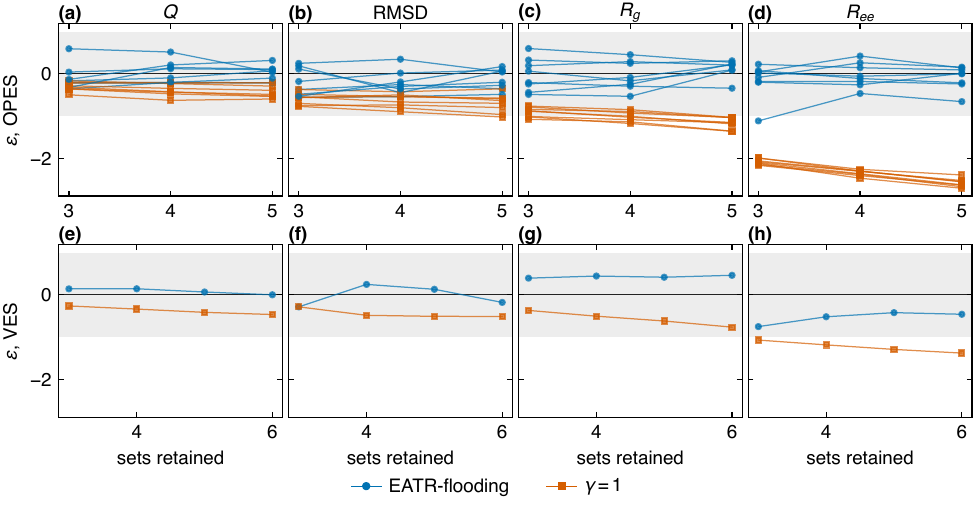}
\caption{Error against the number of lowest-acceleration levels retained. Circles show
EATR-flooding and squares the naive estimate, $\gamma = 1$ (mean of the per-level estimates). Panels (a--d) show OPES, with
one curve per pace and 50 runs per level, and (e--h) show VES, with 200 runs per cutoff.}\label{fig:s_sets}
\end{figure}

Figure~\ref{fig:s_sets} tests whether the inferred rate changes as stronger levels are
added. Levels are ordered by measured acceleration and the $n$ lowest are retained. A
stable plateau supports the result; drift or curvature identifies a ladder that has not
yet constrained $\gamma$ or has begun to overfill the barrier.

\section{Accuracy over ten ln units of rate}\label{sec:s_temperature}
The 312~K benchmark has a $5.95~\kB T$ barrier and usable boosts below about 150. Lowering
the temperature raises the barrier without changing the system; at 270~K it is
$16.09~\kB T$, implying a rate ten ln units slower and an acceleration of nearly $10^6$. This
scan tests whether the estimator remains accurate far beyond the easy benchmark regime.

\begin{figure}[H]\centering\includegraphics[width=\textwidth]{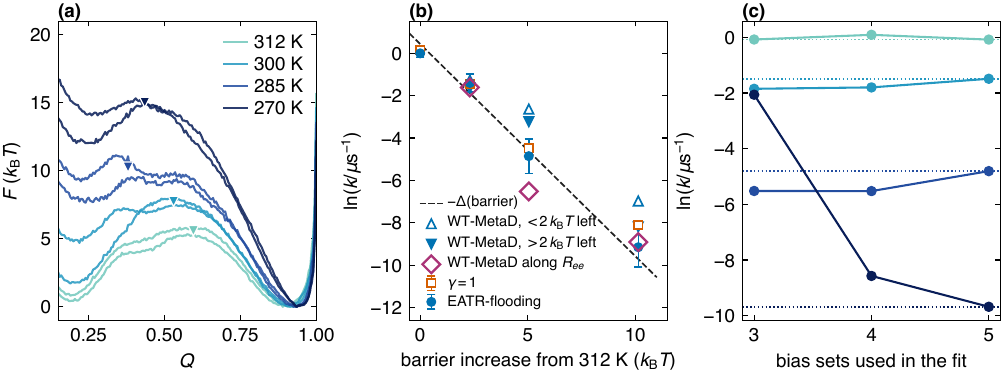}
\caption{Rates across the temperature scan. (a) Free energy along $Q$ from the
well-tempered scan, obtained by reweighting the second half of each run and zeroed at the folded minimum; both
independent runs are drawn and the triangle marks the top of the mean profile, before the
6\% correction described in the text.
(b) Rate against the barrier change from 312~K. The dashed line is the barrier prediction,
$\Delta \ln k = -\Delta(\text{barrier})$, with no free parameter. Filled circles show
EATR-flooding on the five fill levels, as means of bootstrap resamples with their standard
deviations, and open squares show the
same data with $\gamma = 1$. Triangles show EATR on a single WT-MetaD condition per
temperature, open where the achieved residual barrier is below $2~\kB T$ and
filled where it is above; the purple diamonds are the same test along $R_{ee}$ (200 runs per temperature), whose naive estimates, at $-5.6$, $-11.8$ and $-16.4$, are not drawn. (c) The same flooding fits repeated using three, four and five bias
sets; dotted lines mark the five-set value. Panel (c) shows fits to all runs without
resampling, which differ from the bootstrap means of panel (b) by up to 0.5 ln units (at 270~K).}\label{fig:s_temperature}
\end{figure}

At 270, 285, 300 and 312~K, two independent 2~$\mu$s WT-MetaD runs provide reweighted free
energies along $Q$ (Fig.~\ref{fig:s_temperature}a). The 312~K estimate is anchored to the
unbiased $5.95~\kB T$ barrier; because the reweighted estimate remains 6\% low after
2~$\mu$s, all temperatures use the same multiplicative correction. Each temperature then
uses five OPES-flooding levels with 50 runs per level, placed at the same residual barriers
as the 312~K ladder.

Because the unbiased rate is known only at 312~K, the accuracy target is the parameter-free
change $\Delta\ln k=-\Delta(\text{barrier})$. Over the three lower temperatures, which span 7.8 ln units of predicted rate,
EATR-flooding follows that prediction with slope 0.97 and mean offset $+0.26\pm0.44$ ln
units; with 312~K included the slope is 0.93 and the offset $+0.09\pm0.50$. The naive correction has slope 0.83
and reaches an offset of $+1.60$ at 270~K (Fig.~\ref{fig:s_temperature}b). At 312~K, the
EATR-flooding estimate is 0.44 ln units below the independently known reference, within the
one-ln-unit accuracy target despite omitting the excluded region used in the main text.

Accuracy is also insensitive to the achieved fill within each ladder. Measured residual
barriers span $+3.3$ to $-0.6~\kB T$, including levels that pass the barrier top, yet the
per-level rates scatter by only 0.70--1.08 ln units with no systematic trend against fill.

Table~\ref{tab:s18} compares cost with the mean first-passage time implied by the barrier.
The five levels cost 6--14~$\mu$s at every temperature while the physical first-passage
time grows from 0.65~$\mu$s to 16~ms. The gain over unbiased sampling therefore rises from
none at 312~K to roughly 3000-fold at 270~K without losing rate accuracy.
\begin{table}[H]\centering
\caption{Cost of the low-temperature conditions against the mean first-passage time (MFPT) the barrier implies, $e^{\Delta(\text{barrier})}$ times the 312~K value of $0.65~\mu$s. The five OPES-flooding levels carry 50 runs each; the well-tempered condition is the one per temperature that kept more than $2~\kB T$ of barrier, 50 runs at 300 and 285~K and 200 at 270~K. The same protocol along $R_{ee}$ is listed at each temperature, with 200 runs. The boost is the first-passage time over the mean biased escape time. Errors $\varepsilon$ are against the barrier prediction $\Delta\ln k = -\Delta(\text{barrier})$ through the 312~K reference, for the $\gamma$-fitting estimator (EATR-flooding, EATR) and the naive one ($\gamma = 1$).}\label{tab:s18}
\resizebox{\textwidth}{!}{\begin{tabular}{llrrrrrrr}\toprule
$T$ (K) & protocol & runs & cost ($\mu$s) & MFPT ($\mu$s) & cost/MFPT & boost & $\varepsilon$ ($\gamma$ fitted) & $\varepsilon$ ($\gamma = 1$)\\ \midrule
312 & OPES, five levels & 250 & 10.8 & 0.65 & 16.61 & 15 & -0.44 & -0.29 \\
300 & OPES, five levels & 250 & 6.0 & 6.55 & 0.91 & 274 & 0.47 & 0.43 \\
300 & WT-MetaD, one condition & 50 & 17.3 & 6.55 & 2.65 & 19 & 0.26 & 0.16 \\
300 & WT-MetaD, one condition, $R_{ee}$ & 200 & 118.2 & 6.55 & 18.05 & 11 & 0.26 & -3.74 \\
285 & OPES, five levels & 250 & 13.8 & 101 & 0.14 & 1835 & -0.24 & 0.13 \\
285 & WT-MetaD, one condition & 50 & 28.7 & 101 & 0.28 & 177 & 1.38 & -0.77 \\
285 & WT-MetaD, one condition, $R_{ee}$ & 200 & 164.3 & 101 & 1.62 & 123 & -1.90 & -7.23 \\
270 & OPES, five levels & 250 & 5.7 & 16465 & $3.5\times10^{-4}$ & 723976 & 0.56 & 1.60 \\
270 & WT-MetaD, one condition & 200 & 153.4 & 16465 & $9.3\times10^{-3}$ & 21469 & 1.52 & -0.30 \\
270 & WT-MetaD, one condition, $R_{ee}$ & 200 & 123.5 & 16465 & $7.5\times10^{-3}$ & 26674 & 0.80 & -6.69\\ \bottomrule
\end{tabular}}
\end{table}

The number of levels required for accuracy does grow with acceleration
(Fig.~\ref{fig:s_temperature}c). Three levels suffice at 312~K, but at 270~K they give
$\gamma=0.43$ and a rate 7.6 ln units high; all five are required. Since an error in
$\gamma$ shifts the rate by $(1-\gamma)\ln\alpha$, the ladder must constrain $\gamma$ more
tightly as $\ln\alpha$ grows from about 3 at 312~K to 13 at 270~K.

We perform a single WT-MetaD condition per temperature to test whether one growing-bias fill can replace the OPES ladder.
It can at 300~K but our current tests show it is not as accurate at lower temperatures, whether or not the residual-barrier rule is met. 
With 0.85--1.75~$\kB T$ of barrier left (50 runs per temperature), EATR overshoots the prediction by $+0.6$, $+2.0$ and $+2.7$ ln units at 300, 285 and 270~K. Slowing the fill by 5.7 leaves 2.60--3.74~$\kB T$ and brings 300~K to $+0.26$, but 285~K stays at $+1.38$ (the two free-energy runs there differ by $1.58~\kB T$) and 270~K gives $+1.52$ [$+0.10$, $+2.64$] with 200 runs and $\gamma = 0.84$ [0.74, 0.96]. 
Along $R_{ee}$, with 200 runs per temperature and hill heights chosen for one predicted escape time ($h_0 = 1.1$, 1.9 and 4.0~kJ/mol), the errors are $+0.26$ [$-0.19$, $+0.64$], $-1.90$ [$-3.01$, $-0.96$] and $+0.80$ [$-0.85$, $+2.19$] at 300, 285 and 270~K, with $\gamma = 0.40$ [0.34, 0.45], 0.62 [0.55, 0.69] and 0.61 [0.53, 0.69]. The naive estimate there is 3.7, 7.2 and 6.7 ln units low. The 285~K interval excludes the barrier prediction.
The limit is the resolution of $\gamma$: at 270~K, where $\ln\alpha \simeq 10$, each 0.1 of $\gamma$ is one ln unit of rate, and one WT-MetaD condition leaves a 95\% interval of about $\pm1.3$ ln units, comparable to the five-level ladder's ($\pm1.6$) at 27 times the cost.
A first fit of the 270~K condition with 50 runs shows that it gave an error of $+0.03$ only because $\gamma$ sat near its bound of 1 ([0.83, 1.00]), and the error rose from $+0.27$ to $+1.45$ as subsets grew from 5 to 80 runs and the fit resolved $\gamma$ below 1. This is Fig.~\ref{fig:s_temperature}(c) restated for a growing bias: the number of independent fills constraining our estimate of $\gamma$ must grow with the acceleration.

\section{When fitting $\gamma$ reduces accuracy}
\begin{figure}[H]\centering\includegraphics[width=\textwidth]{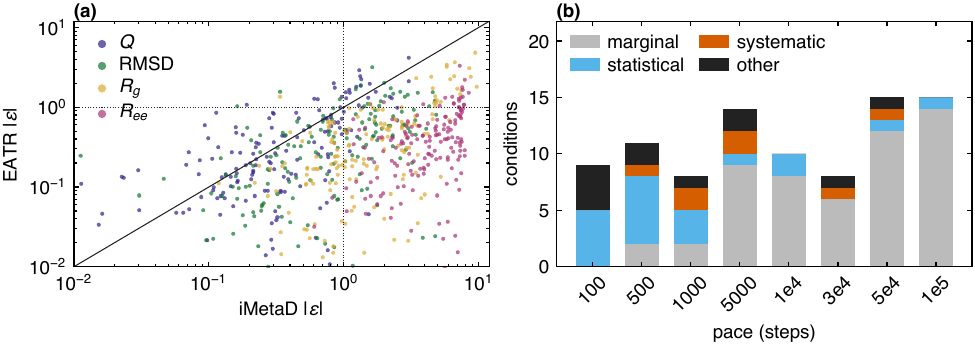}
\caption{(a) Absolute error of both estimators for every full fit of the grid;
above the diagonal EATR is worse. (b) Conditions where EATR is worse are counted by pace and class.
The classes are marginal (both within one ln unit, differing by less than 0.3), statistical (reference
inside the EATR interval, or difference below the subset spread), systematic (EATR outside
one ln unit, iMetaD inside) and other.}\label{fig:s_worse}
\end{figure}

Figure~\ref{fig:s_worse} classifies the 90 of 639 full-grid fits in which EATR is less
accurate than iMetaD. Most differences are marginal or consistent with sampling noise;
11 place EATR outside one ln unit while iMetaD remains inside, and only 7 of those are systematic rather than statistical. In these systematic
cases EATR overestimates the rate, by 1.0 to 1.6 ln units, with a low fitted $\gamma$
(0.32--0.67); six are for $Q$ or RMSD and five are at bias factor 2.

\section{Validity checks for VES cutoff flooding}
\begin{figure}[H]\centering\includegraphics[width=\textwidth]{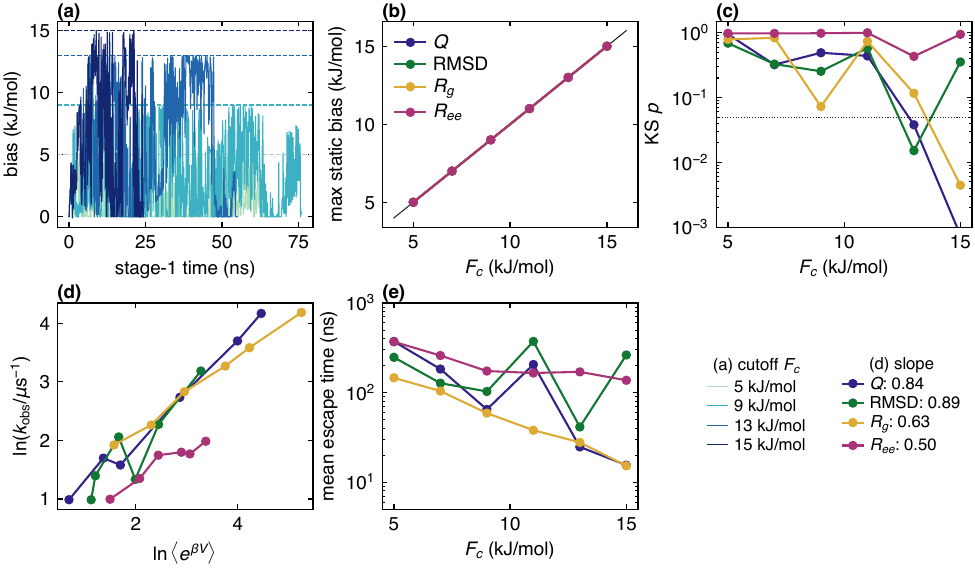}
\caption{Checks of VES cutoff flooding. (a) Stage-1 bias at the walker for $Q$, with the cutoff shown as a dashed line.
(b) Maximum of the static stage-2 bias against the cutoff. (c) KS $p$ of the escape times
against an exponential for 200 runs per cutoff, with the dotted line at 0.05. (d) Observed rate against
acceleration, with the least-squares slope. (e) Mean escape time per
cutoff.}\label{fig:s_ves}
\end{figure}

Figure~\ref{fig:s_ves} checks each step in the VES analysis. The optimized bias reaches its
cutoff, and the static production bias has maximum $F_c$ in all 24 sets. Escape times are
exponential except at the highest cutoffs for $Q$, RMSD and $R_g$; the slope of
$\ln k_{\rm obs}$ against acceleration agrees with fitted $\gamma$ except for $R_{ee}$.
For $Q$ and RMSD, non-monotonic escape times reveal that the single stage-1 optimization
was stopped before high-cutoff biases converged. EATR-flooding remains usable because it
orders levels by measured acceleration rather than nominal cutoff.

\section{Assessing whether iMetaD data can be used within EATR-flooding}
In our previous study, we tested using time-dependent biasing within our EATR-flooding framework, which apparently was allowable.\footnote{N. Mazzaferro, W. J. Pe\~{n}a Ccoa, P. Cossio and G. M. Hocky, J. Chem. Theory Comput. \textbf{22}, 7570 (2026).}
With our larger data set, we tested this general strategy. 
Table~\ref{tab:s8} treats eight deposition paces at fixed bias factor and height as if they
were static flooding levels. This construction is inaccurate for each CV in this study. 
The slope of $\ln k_{\rm obs}$ against whole-trajectory acceleration never approaches 1, and the rate
misses by as much as 2.2 ln units. In general, varying a set of different paces or heights does not appear to result in one shared
static relationship between bias and rate.
\begin{table}[H]\centering
\caption{EATR-flooding across the eight deposition paces of a growing bias, at each fixed bias factor and height: range over those series of the slope of $\ln k_{\rm obs}$ against the whole-trajectory acceleration, of the fitted $\gamma$, and the largest error. Every series individually is in the analysis repository.}\label{tab:s8}
\begin{tabular}{lcccc}\toprule
CV & series & slope & $\gamma$ & largest $\varepsilon$\\ \midrule
$Q$ & 16 & 0.31--0.49 & 0.34--0.55 & +1.81 \\
RMSD & 16 & 0.22--0.36 & 0.30--0.47 & +1.67 \\
$R_g$ & 8 & 0.16--0.25 & 0.27--0.39 & +2.19 \\
$R_{ee}$ & 14 & 0.26--0.45 & 0.31--0.50 & -2.20\\ \bottomrule
\end{tabular}
\end{table}

\end{document}